\documentclass[1p,fleqn,times]{elsarticle}

\usepackage{hyperref}

\usepackage{geometry} 

\usepackage{natbib}
\usepackage{amsmath,amssymb}
\usepackage{mathrsfs}
\usepackage{physics}
\usepackage{xcolor}
\usepackage{mathtools}
\usepackage{graphicx}
\usepackage{rotating}
\usepackage{blindtext}
\usepackage{multirow}
\usepackage{subcaption}
\usepackage{bibentry}

\newcommand{\Vbar}{\overline{\mathbf{V}}^e}

\newcommand{\dev}[1]{\mbox{dev}\left({#1}\right)}

\newcommand{\normal}{\mathbf{\hat{n}}}

\newcommand{\reflected}{\mbox{\scriptsize{ref}}}
\newcommand{\constant}{\mbox{\scriptsize{const}}}

\graphicspath{{} {Images/}}

\makeatletter
\def\ps@pprintTitle{%
  \let\@oddhead\@empty
  \let\@evenhead\@empty
  \def\@oddfoot{}%
  \let\@evenfoot\@oddfoot}
\makeatletter
\def\ps@pprintTitle{%
  \let\@oddhead\@empty
  \let\@evenhead\@empty
  \def\@oddfoot{}%
  \let\@evenfoot\@oddfoot}
\makeatother
\usepackage{fancyhdr}

\fancypagestyle{pprintTitle}{
\lhead{}\chead{}\rhead{}\lfoot{\textcopyright \small ~British Crown Owned Copyright 2021/AWE}\cfoot{\thepage}\rfoot{}
}

\begin{document}

\begin{frontmatter}

\title{A unified diffuse interface method for the interaction of rigid bodies with elastoplastic solids and multi-phase mixtures}

\author[mymainaddress]{Tim Wallis \corref{mycorrespondingauthor}}
\cortext[mycorrespondingauthor]{Corresponding author}
\ead{tnmw2@cam.ac.uk}

\author[mysecondaryaddress]{Philip T. Barton}
\author[mymainaddress]{Nikolaos Nikiforakis}

\address[mymainaddress]{Department of Physics, University of Cambridge, Cavendish Laboratory, JJ Thomson Avenue, CB3 0HE, UK}
\address[mysecondaryaddress]{AWE Aldermaston, Reading, Berkshire, RG7 4PR, UK}

\begin{abstract}

This work outlines a new multi-physics-compatible immersed rigid body method for Eulerian finite-volume simulations. To achieve this, rigid bodies are represented as a diffuse scalar field and an interface seeding method is employed to mediate the interface boundary conditions.
The method is based on an existing multi-material diffuse interface method that is capable of handling an arbitrary mixture of fluids and elastoplastic solids. The underlying method is general and can be extended to a range of different applications including high-strain rate deformation in elastoplastic solids and reactive fluid mixtures. As such, the new method presented here is thoroughly tested through a variety of problems, including fluid-rigid body interaction, elastoplastic-rigid body interaction, and detonation-structure interaction. Comparison is drawn between both experimental work and previous numerical results, with excellent agreement in both cases.
The new method is straightforward to implement, highly local, and parallelisable. This allows the method to be employed in three dimensions with multiple levels of adaptive mesh refinement using complex immersed geometries. The rigid body field can be static or dynamic, with the THINC interface reconstruction method being used to keep the interface sharp in the dynamic case.

\textcopyright ~British Crown Owned Copyright 2021/AWE
\end{abstract}


\end{frontmatter}

\pagestyle{pprintTitle}
\thispagestyle{pprintTitle}

\section{Introduction}
Immersed rigid boundaries feature in the simulation of a wide variety of different physical applications. These range from fluid-solid interaction problems to elastoplastic machining studies. The inclusion of a rigid body in a simulation is most often to save computational expense; discretising inside a solid body and solving a fully elastoplastic system of equations is expensive for a number of reasons. In certain applications, the evolution inside a solid body may not be relevant, and in this case it saves time and expense to replace the solid with a rigid boundary.

This paper introduces a new diffuse interface method for the inclusion of static and dynamic rigid bodies in Cartesian Eulerian finite-volume simulations. A variety of other methods exist to model rigid boundaries, designed for a range of different problems including inviscid hypersonic flows, viscous Navier-Stokes problems and turbulence modelling. Many fluid-structure interaction studies consider body-fitted or unstructured meshes. These methods, however, often feature some difficulty and expense in creating suitable meshes around complex geometries in three dimensions. Cartesian approaches also exist, offering a more straightforward approach to mesh generation and the data structures that describe them. Such approaches include level-set based Ghost Fluid methods such as \citet{UdaykumarRigid} or cut-cell approaches such as \citet{NandanCutCell}. However these methods have their own associated problems. Recently \citet{DumbserRigid} also include rigid bodies with the use of a diffuse interface model, incorporating the effect of the rigid body as a forcing term in the momentum equation, using the ADER-WENO approach.

This work chooses a diffuse interface representation for both the material interfaces and rigid body interfaces. The methods presented build on the work of \citet{WallisFluxEnriched}, where novel flux-modifiers and interface seeding routines were employed to mediate void-opening, slip and fracture internal boundary conditions in a diffuse interface context, without the use of level sets. The work at hand follows a very similar approach by adding an additional scalar field $r$, the rigid body volume fraction, into the system of equations to track rigid bodies. From here, an analogous interface seeding method is derived to enable rigid boundaries to be incorporated in the system. The rigid body volume fraction is treated as any other volume fraction, allowing the THINC reconstruction methods of \citet{XiaoTHINC} to be used to keep interfaces sharp. 

The work of \citet{WallisFluxEnriched} used a reduced-equation Allaire-type \cite{Allaire} multi-material diffuse interface scheme, adapted by \citet{Barton2019} to include elastoplastic solids. On top of this underlying multi-material method, additional history parameters such as plastic strain, and scalar fields such as the void volume fraction, were then used to include additional physics in a straightforward manner. This method features a separate equation of state for each material, which then combine in mixture regions according to thermodynamically consistent rules, weighted by the volume fraction they occupy. This is in contrast to other mixture methods such as multi-fluid mixing, as presented by \citet{BatesRM} and \citet{MosedaleLES}, although it can be noted that similar numerical methods may be used in both cases.

The method at hand forms part of a rich history of research into pure-Eulerian Godunov methods for coupled solid-fluid type problems. This background includes the development of the underpinning hyperbolic hyperelastic models of elastoplastic solids \cite{romenski:1989,plohr:1992,godunov:2003,peshkov:2014}, the development of single material Godunov numerical methods \cite{wang:1993,miller:2001,titarev:2008,barton:2009,barton:2010a,hill:2010}, the extension to multi-materials using sharp interface methods \cite{miller:2002,barton:2010b,BartonLevelSet,barton:2013,schoch:2013,lopez:2014}, the development of diffuse interface methods \cite{favrie:2009,favrie:2012,hank:2017,Barton2019}, and the development of the models to include additional physics such as damage and reactive materials in \cite{michael:2018,barton:2016,barton:2017,WallisMultiPhysics,WallisFluxEnriched}. The references presented here are by no means exhaustive, but rather they represent a selection of seminal works from an area which remains very active.

As outlined by \citet{WallisFluxEnriched}, the major benefit of the new scheme is its straightforward, practical approach. Very little additional numerical machinery must be added to incorporate rigid bodies, and any desired physics. For the hyperbolic update, the same set of partial differential equations is solved across the entire domain, incorporating the effect of all materials in the mixture. The numerical methods required to mediate the rigid boundary conditions are highly local to the rigid body interface, avoiding the parallel communication that can be required for other methods. This simplicity allows the method to parallelise efficiently over several layers of adaptive mesh refinement (AMR) and in three dimensions. Additionally, as rigid bodies are represented as a continuous scalar field in a standard Cartesian mesh, no additional effort must be made in creating an appropriate mesh, or re-meshing, and the use of body fitted meshes is avoided. 
The new methods developed in this work are entirely compatible with the previous works on which it is based, as the same underlying multi-material system of equations employed by \citet{WallisFluxEnriched} is used here. This allows the new rigid body method to be applied to a wide variety of applications, including single material fluid flow, elastoplastic-rigid interaction, and reactive fluid mixtures (as presented by \citet{WallisMultiPhysics}). This work will therefore present a number of validation cases covering this range, with both experimental and numerical comparison.

\section{Governing Theory}

The system considered in this work will be laid out as follows. Firstly, the full multi-material base system of equations and its thermodynamics will be briefly outlined. The base system has been laid out in detail in previous works \cite{Barton2019, WallisFluxEnriched, WallisMultiPhysics} and is not affected by the later additions of a rigid body, so shall not be repeated here in depth. Subsequently, the additional components required for the rigid body will be introduced. From here, a smaller subsystem will be derived, for use in single material fluid-structure interaction problems, which will form a large part of this work. 

\subsection{Base System}
\subsubsection{Evolution Equations}

The base system, devised by \citet{Barton2019}, is an Allaire-type \cite{Allaire} multi-material diffuse interface system, capable of accounting for an arbitrary mixture of both fluids and elastoplastic solids. The state of any material $l$ is characterised by its phasic density $\rho_{(l)}$, volume fraction $\phi_{(l)}$, symmetric left unimodular stretch tensor $\Vbar$, velocity vector $\mathbf{u}$, and internal energy $\mathscr{E}$. The simplifying assumption is made that materials in a mixture are in mechanical and thermal equilibrium, resulting in a reduced equation system where materials share the same momentum, energy and deformation equations.
For $l=1,\ldots, N$ materials:
\begin{align}
\frac{\partial \phi_{(l)}}{\partial t} + \frac{\partial \phi_{(l)} u_k}{\partial x_k} &= \phi_{(l)}\frac{\partial u_k}{\partial x_k}\\
\frac{\partial \phi_{(l)}\rho_{(l)}}{\partial t} + \frac{\partial \phi_{(l)}\rho_{(l)} u_k}{\partial x_k} &= 0 \\
\frac{\partial \rho u_i }{\partial t} + \frac{\partial (\rho u_i u_k -\sigma_{ik}) }{\partial x_k} &= 0\\
\frac{\partial \rho E }{\partial t} + \frac{\partial (\rho E u_k - u_i \sigma_{ik}) }{\partial x_k} &= 0\\
\frac{\partial \Vbar_{ij} }{\partial t} + \frac{\partial \left( \Vbar_{ij}  u_k - \Vbar_{kj} u_i \right) }{\partial x_k} &= \frac{2}{3}\Vbar_{ij}\frac{\partial u_k}{\partial x_k} - u_i\beta_j - \boldsymbol\Phi_{ij} \ , 
\end{align}
where $E=\mathscr{E}+|\mathbf{u}|^2/2$ denotes the specific total energy, $\boldsymbol\sigma$ denotes the Cauchy stress tensor, $\beta_j=\partial\Vbar_{kj}/\partial x_k$, and $\boldsymbol\Phi$ represents the contribution from plastic effects. The plastic source term $\boldsymbol\Phi$ limits the possible range of elastic deformation and initiates the growth of plastic deformation. This work follows the method of convex potentials, defining this source term as 
\begin{equation}
 \boldsymbol\Phi = \chi\pdv{\varphi(\boldsymbol\sigma)}{\boldsymbol\sigma}\Vbar \ ,
\end{equation}
where $\chi$ is a plastic multiplier and $\varphi$ is a convex potential, both of which are closure models. Further details of this source term can be found in \citet{Barton2019} and comprehensive details of numerous alternative closure models can be found in \citet{asaro:2006}.

Some multi-physics closure models introduce a dependence on material history variables. For example, the equivalent plastic strain $\varepsilon_{p(l)}$ or the damage parameter $D_{(l)}$ in the case of fracture \cite{WallisFluxEnriched}, or the reaction progress variable $\lambda$ for reactive fluid mixtures \cite{WallisMultiPhysics}. 
For these variables, additional evolution equations are required:
\begin{eqnarray}
\frac{\partial \rho_{(l)}\phi_{(l)} \alpha_{(l)}}{\partial t} + \frac{\partial \rho_{(l)} \phi_{(l)} \alpha_{(l)} u_k}{\partial x_k} &=& \rho_{(l)}\phi_{(l)}\dot{\alpha}_{(l)} \ .
\end{eqnarray}
Here $\alpha_{(l)}$ represents any such history parameter which is advected and evolved with a material as time progresses. A material may have more than one history variable, in which case $\alpha$ represents a vector.

\subsubsection{Thermodynamics}
\label{sec:thermodynamics}
The internal energy $\mathscr{E}$ for each material is defined by an equation-of-state that conforms to the general form:
\begin{equation}\label{eq_eos_gen}
\mathscr{E}_{(l)}\left(\rho_{(l)},T_{(l)},\dev{\mathbf{H}^e},\alpha_{(l)}\right) = \mathscr{E}_{(l)}^c\left(\rho_{(l)},\alpha_{(l)}\right)+ \mathscr{E}_{(l)}^t\left(\rho_{(l)},T_{(l)}\right) + \mathscr{E}_{(l)}^s\left(\rho_{(l)},\dev{\mathbf{H}^e},\alpha_{(l)}\right) \ ,
\end{equation}
where
\begin{equation}
\dev{\mathbf{H}^e} = \ln\left(\Vbar\right) \ 
\end{equation}
is the deviatoric Hencky strain tensor and $T$ is the temperature. The three terms on the right hand side are the contribution due to cold compression or dilation, $\mathscr{E}_{(l)}^c\left(\rho_{(l)},\alpha_{(l)}\right)$, the contribution due to temperature deviations, $\mathscr{E}_{(l)}^t\left(\rho_{(l)},T_{(l)}\right)$, and the contribution due to shear strain $\mathscr{E}_{(l)}^s\left(\rho_{(l)},\dev{\mathbf{H}^e},\alpha_{(l)}\right)$. The cold compression energy will generally be provided by the specific closure model for each material. The thermal energy is given by
\begin{eqnarray}
\mathscr{E}_{(l)}^t(\rho_{(l)},T)&=& C_{(l)}^{\text{V}}\left(T-T_{(l)}^0\theta_{(l)}^D\left(\rho_{(l)}\right)\right) \ ,
\end{eqnarray}
where $C_{(l)}^{\text{V}}$ is the heat capacity, $T_{(l)}^0$ is a reference temperature, and $\theta_{(l)}^{\text{D}}(\rho_{(l)})$ is the non-dimensional Debye temperature.
The Debye temperature is related to the Gr\"uneisen function, $\Gamma(\rho_{(l)})$, via
\begin{equation}
\Gamma_{(l)}(\rho_{(l)}) = \frac{\partial \ln\theta_{(l)}^{\text{D}}(\rho_{(l)})}{\partial \ln(1/\rho_{(l)})} = \frac{\rho_{(l)}}{\theta_{(l)}^{\text{D}}(\rho_{(l)})}\frac{\partial \theta_{(l)}^{\text{D}}(\rho_{(l)})}{\partial\rho_{(l)}} \ .
\end{equation}
The specific form of the Gr\"uneisen function for each material must be provided, but it is generally taken to be constant. The elastic shear energy is given by
\begin{equation}
\mathscr{E}_{(l)}^s(\rho_{(l)},\dev{\mathbf{H}^e},\alpha_{(l)}) =  \frac{G_{(l)}\left(\rho_{(l)},\alpha_{(l)}\right)}{\rho_{(l)}}  \mathcal{J}^2\left(\dev{\mathbf{H}^e}\right) \ ,
\end{equation}
where $G_{(l)}(\rho_{(l)},\alpha_{(l)})$ is the shear modulus, and
\begin{equation}
\mathcal{J}^2(\dev{\mathbf{H}^e}) = \tr\left(\dev{\mathbf{H}^e}\dev{\mathbf{H}^e}^{\text{T}}\right)
\end{equation}
is the second invariant of shear strain.

For each component, the Cauchy stress, $\boldsymbol\sigma$, and pressure, $p$, are inferred from the second law of thermodynamics and classical arguments for irreversible elastic deformations:
\begin{eqnarray}
\boldsymbol\sigma_{(l)} &=& p_{(l)}\mathbf{I} + \dev{\boldsymbol\sigma_{(l)}} \\
p_{(l)} &=& \rho^{2}_{(l)} \frac{\partial\mathscr{E}_{(l)}}{\partial \rho_{(l)}} \label{eq:p_energy_derivative}\\ 
\dev{\boldsymbol\sigma_{(l)}} &=& 2G_{(l)}\dev{\mathbf{H}^e} \ .
\end{eqnarray}
Although it might appear that the model describes solid materials, fluids can be considered a special case where the shear modulus is zero, resulting in a spherical stress tensor and no shear energy contribution. This formulation lends itself well to diffuse interface modelling where different phases that share the same underlying model can combine consistently in mixture regions.

It can be seen that, by presenting the internal energy in the form outlined, equation \eqref{eq:p_energy_derivative} can be written in the form:
\begin{align}
 p_{(l)} = p_{\text{\scriptsize{ref}},(l)} + \rho_{(l)}\Gamma_{(l)}\left(\mathscr{E}_{(l)} - \mathscr{E}_{\text{\scriptsize{ref}},(l)}\right) \label{eq:MieGruneisenEOS} \ .
\end{align}
Here, $\mathscr{E}_{\text{\scriptsize{ref}},(l)} = \mathscr{E}^c_{(l)} + \mathscr{E}^s_{(l)}$ and $p_{\text{\scriptsize{ref}},(l)} = \rho^{2}_{(l)} \frac{\partial\mathscr{E}_{\text{\scriptsize{ref}},(l)}}{\partial \rho_{(l)}}$. This is the form of the standard Mie-Gr\"uneisen equation-of-state, and it is therefore the choice of the reference curves $p_{\text{\scriptsize{ref}}}, \mathscr{E}_{\text{\scriptsize{ref}}}$ and $\Gamma$ which will convey specific material properties. This form is not limited to solely fluids or solids, with the choice of reference curves capable of incorporating cold-compression, thermal, and shear effects. A variety of different materials will be considered in the validation tests, which are detailed in Section \ref{sec:ClosureModels}.

Following the approach of \citet{Barton2019}, rather than calculating the full acoustic tensor, the sound-speed $c_{(l)}$ for a material is approximated by:
\begin{equation}
 c_{(l)}^2 \approx a_{(l)}^2 + \frac{4}{3} b_{(l)}^2 \ ,
\end{equation}
where
\begin{equation}
 a_{(l)}^2 = \left(\pdv{p_{(l)}}{\rho_{(l)}}\right)_S =  \frac{\Gamma_{(l)} p_{(l)}}{\rho_{(l)}} + \left(p_{(l)}-p_{\text{\scriptsize{ref}},(l)}\right)\left(\frac{1}{\rho_{(l)}} + \pdv{\Gamma_{(l)}}{\rho_{(l)}}\right)+\left(\pdv{p_{\text{\scriptsize{ref}},(l)}}{\rho_{(l)}}-\Gamma_{(l)}\rho_{(l)}\pdv{\mathscr{E}_{\text{\scriptsize{ref}},(l)}}{\rho_{(l)}}\right) \ ,
\end{equation}
where $S$ is the entropy and
\begin{equation}
 b_{(l)}^2 = \frac{G_{(l)}}{\rho_{(l)}} \ 
\end{equation}
is the transverse wave speed squared.

\subsubsection{Closure Models}
\label{sec:ClosureModels}

For the case of fluids like air, a simple ideal gas law will be used. In this case, $p_{\text{\scriptsize{ref}}} = 0, \mathscr{E}_{\text{\scriptsize{ref}}} = 0$, and the material is entirely specified by the choice of the adiabatic index $\gamma = \Gamma + 1$ which is taken to be constant. This reduces the form of Equation \ref{eq:MieGruneisenEOS} to:
\begin{align}
 p_{(l)} = (\gamma_{(l)}-1)\rho_{(l)} \mathscr{E}_{(l)} \ .
\end{align}

When elastoplastic solids are employed, this work follows previous studies by using the equation of state from \citet{RomenskiiEOS}. For this equation of state, $\Gamma(\rho)=\Gamma_0$, where $\Gamma_0$ is a material dependent constant, and:
\begin{align} \label{eq:Romenskii}
\mathscr{E}_{\text{\scriptsize{ref}},(l)}(\rho_{(l)}) &= \mathscr{E}_{(l)}^c + \mathscr{E}^s_{(l)} = \frac{K_{0,(l)}}{2\rho_{(l)}\alpha_{(l)}^2}\left(\left(\frac{\rho_{(l)}}{\rho_{0,(l)}}\right)^{\alpha_{(l)}}-1\right)^2 + \frac{G_{(l)}\left(\rho_{(l)},\alpha_{(l)}\right)}{\rho_{(l)}}  \mathcal{J}^2\left(\dev{\mathbf{H}^e}\right) \\
G_{(l)}(\rho_{(l)}) &= G_{0,(l)}\left(\frac{\rho_{(l)}}{\rho_{0,(l)}}\right)^{\beta_{(l)}+1} \\
p_{\text{\scriptsize{ref}},(l)}(\rho_{(l)}) &= \rho^{2}_{(l)} \frac{\partial\mathscr{E}_{\text{\scriptsize{ref}},(l)}}{\partial \rho_{(l)}} \ .
\end{align}

Finally, when reactive fluids are considered, the Jones-Wilkins-Lee (JWL) equation of state is employed for both the reactants and products of the explosive. For this equation of state $\Gamma(\rho)=\Gamma_0$ and 
\begin{align}
  \mathscr{E}_{\text{\scriptsize{ref}},(l)}(\rho_{(l)}) &= \frac{{\cal A}_{(l)}}{{\cal R}_{1,(l)}\rho_{0,(l)}}e^{-{\cal R}_{1,(l)}\frac{\rho_{0,(l)}}{\rho_{(l)}}}+\frac{{\cal B}_{(l)}}{{\cal R}_{2,(l)}\rho_{0,(l)}}e^{-{\cal R}_{2,(l)}\frac{\rho_{0,(l)}}{\rho_{(l)}}} \\ 
  p_{\text{\scriptsize{ref}},(l)}(\rho_{(l)}) &= \rho^{2}_{(l)} \frac{\partial\mathscr{E}_{\text{\scriptsize{ref}},(l)}}{\partial \rho_{(l)}} = {\cal A}_{(l)}e^{-{\cal R}_{1,(l)}\frac{\rho_{0,(l)}}{\rho_{(l)}}}+{\cal B}_{(l)}e^{-{\cal R}_{2,(l)}\frac{\rho_{0,(l)}}{\rho_{(l)}}}
\end{align}

\subsection{Mixture Rules}

Having defined the thermodynamics and closure models for each separate material, they may now be combined into a mixture. In order to do this, mixture rules must be provided to represent the state of regions containing multiple materials in a thermodynamically consistent way. The following mixture rules are applied \cite{Allaire, Barton2019}:
\begin{eqnarray}
1 &=& \sum_{l=1}^N\phi_{(l)}\label{vof_mix_rule} \\
\rho &=& \sum_{l=1}^N \phi_{(l)} \rho_{(l)} \\
\rho \mathscr{E} &=& \sum_{l=1}^N \phi_{(l)} \rho_{(l)} \mathscr{E}_{(l)}\label{eq:energyMixtureRule}\\ 
G &=& \frac{\sum_{l=1}^N\left(  \phi_{(l)} G_{(l)}(\rho_{(l)},\alpha_{(l)})/\Gamma_{(l)}\right)}{\sum_{l=1}^N\left(\phi_{(l)}/\Gamma_{(l)}\right)}\label{eq:ShearMixtureRule}\\
c^2 &=& \frac{\sum_{l=1}^N \left( \phi_{(l)} Y_{(l)} c_{(l)}^2 / \Gamma_{(l)} \right)}{\sum_{l=1}^N\left(\phi_{(l)}/\Gamma_{(l)}\right)} \ ,
\end{eqnarray}
where $c$ is the sound-speed and $Y_{(l)} = ({\phi_{(l)}\rho_{(l)}})/{\rho} $ is the mass fraction.

Using these mixture rules and the form of the Mie-Gr\"uneisen equation of state, the total hydrodynamic pressure and total stress can be obtained from the total energy. Initially, to calculate the total pressure, equation \ref{eq:energyMixtureRule} is combined with the isobaric assumption that forms part of the mechanical equilibrium of the underlying \citet{Allaire} model. First the internal energy is obtained:
\begin{equation}
\rho \mathscr{E} = \rho E - \frac{1}{2}\rho|\mathbf{u}|^2 \ .
\end{equation}
Then the Mie-Gr\"uneisen form of the equation of state is used to rewrite equation \ref{eq:energyMixtureRule}, using the isobaric assumption:
\begin{equation}
\rho \mathscr{E} = \sum_{l=1}^N \phi_{(l)} \rho_{(l)} \left(\mathscr{E}_{\text{\scriptsize{ref}},(l)} + \frac{(p - p_{\text{\scriptsize{ref}},(l)})}{\rho_{(l)}\Gamma_{(l)}}\right) \ .
\end{equation}
Finally, this equation can be rewritten in terms of the total pressure $p$:
\begin{equation}
p = \frac{\rho \mathscr{E} - \sum_{l} \phi_{(l)}\left(\rho_{(l)}\mathscr{E}_{\text{\scriptsize{ref}},(l)} - \frac{p_{\text{\scriptsize{ref}},(l)}}{\Gamma_{(l)}}\right)}{\sum_{l}\frac{\phi_{(l)}}{\Gamma_{(l)}}} \ ,
\end{equation}
which may alternatively be written as a weighted sum of component pressures:
\begin{equation}
 p = \frac{\sum_{l=1}^N\left(  \phi_{(l)} p_{(l)}/\Gamma_{(l)}\right)}{\sum_{l=1}^N\left(\phi_{(l)}/\Gamma_{(l)}\right)}
\end{equation}

To obtain the total stress, the total pressure is combined with the mechanical equilibrium assumption that all materials share the same deviatoric strain and the mixture rule for the shear modulus \ref{eq:ShearMixtureRule}:
\begin{equation}
\boldsymbol\sigma = p\mathbf{I} + 2G\dev{\mathbf{H}^e} \\
\end{equation}

The sound-speed mixture rule is used for the estimation of the global time-step and the calculation of the wave speeds required for the flux calculation, as detailed in \citet{Barton2019}.

\subsection{Rigid bodies}

Rigid bodies are represented as scalar fields. This field, $r$, varies from 0 to 1, with the $r=0.5$ contour marking the rigid body interface. When the rigid body is dynamic it evolves under its own velocity field $\mathbf{v}_r$, which can be any required function of space and time, but the velocity does not mutually interact with the flow at large; it is simply imposed on the rigid body. This gives the evolution equation:
\begin{align}\label{eq:revo}
\frac{\partial r}{\partial t} + (\mathbf{v}_r)_k\frac{\partial r }{\partial x_k} &= 0 \ .
\end{align}

\subsection{Reduced System}

Clearly not every one of the equations presented above will be relevant for a given application. Frequently, single-material fluid tests are used to examine the performance of a rigid body method through fluid-solid interaction tests. In this case, a reduced system of equations can be employed, simply consisting of the Euler equations, plus the rigid body evolution law:
\begin{align}
\frac{\partial \rho}{\partial t} + \frac{\partial \rho u_k}{\partial x_k} &= 0\\
\frac{\partial \rho u_i }{\partial t} + \frac{\partial (\rho u_i u_k -\sigma_{ik}) }{\partial x_k} &= 0\\
\frac{\partial \rho E }{\partial t} + \frac{\partial (\rho E u_k - u_i \sigma_{ik}) }{\partial x_k} &= 0 \\
\frac{\partial r}{\partial t} + (\mathbf{v}_r)_k\frac{\partial r }{\partial x_k} &= 0 \ .
\end{align}

Here the stress tensor takes the simple form $\sigma_{ij} = - p \delta_{ij}$, where $p$ is the hydrodynamic pressure and $\delta$ is the Dirac delta function. 

This system is intended to clarify the method in simple fluid-structure interaction problems. However, the full system of equations is still employed when considering more complex situations. For example, the deformation tensor and plastic history variables must also be included alongside this system in elastoplastic-rigid interaction studies.

\section{Numerical Approach}

Apart from the new algorithms required for rigid boundaries, the numerical method for the full multi-material system is the same as presented in \citet{WallisFluxEnriched}. Therefore, the bulk of the method shall not be repeated, but the core algorithms are summarised here.
For clarity, the methods are described in the context of the reduced equation system.

The model is solved on a Cartesian mesh with local resolution adaptation in space and time. This is achieved using the AMReX software from Lawrence Berkeley National Laboratory \cite{amrex}, which includes an implementation of the structured adaptive mesh refinement (SAMR) method of \citet{berger:1988} for solving hyperbolic systems of partial differential equations.
In this approach, cells of identical resolution are grouped into logically rectangular sub-grids or `patches'. 
Refined grids are derived recursively from coarser ones, based upon a flagging criterion, to form a hierarchy of successively embedded levels.
All mesh widths on level $l$ are $r_l$-times finer than on level $l-1$, i.e. $\Delta t_l:=\Delta t_{l-1}/r_l$ and $\Delta \mathbf{x}_{l}:=\Delta \mathbf{x}_{l-1}/r_l$ with $r_l\in\mathbb{N}, r_l\ge 2$ for $l>0$ and $r_0=1$. 
The numerical scheme is applied on level $l$ by calling a single-grid update routine in a loop over all patches constituting the level. 
The discretisation of the constitutive models does not differ between patches or levels, so for clarity the method shall be described for a single sub-grid.
Cell centres are denoted by the indices $i,j,k\in\mathbb{Z}$ and each cell $C^{l}_{ijk}$ has the dimensions $\Delta \mathbf{x}^l_{ijk}$.

It is convenient when describing the numerical method to express the system of equations in compact vector form by separating it into various qualitatively different parts: a conservative hyperbolic part for each spatial dimension and non-conservative terms from the rigid volume fraction update for each spatial dimension:
\begin{equation}
 \frac{\partial \mathbf{U}}{\partial t}+ \frac{\partial \mathbf{F}_k}{\partial x_k} =  \mathbf{s}_{\text{\scriptsize{non-con.}}} \ . \label{eq:sys_vec_form} 
\end{equation}
Note that the non-conservative term will also include the non-conservative updates for the material volume fraction and stretch tensor if the full system is used. For the reduced case, this system is given by:
\begin{align}
 \pdv{t}\mqty(	\rho \\
		\rho u_i \\
		\rho E \\
		r\\)  
		+ \pdv{x_k}\mqty(	\rho u_k \\
					\rho u_iu_k -\sigma_{ik} \\
					\rho Eu_k - u_i\sigma_{ik} \\
					r\mathbf{v}_{r,k} \\)
		= \mqty(0 \\
			0 \\
			0 \\
			r\frac{\partial \mathbf{v}_{r,k} }{\partial x_k} \\) \ .
\end{align}

The inhomogeneous system is integrated for time intervals $[t^n,t^{n+1}]$, where the time-step $\Delta t=t^{n+1}-t^n$ is chosen to be a fraction of the global maximum allowable time step required for stability of the hyperbolic update method \cite{CFL}.

\subsection{Hyperbolic Update}

Replacing the spatial derivatives with a discretised conservative approximation, the hyperbolic system can be written
\begin{equation}\label{sys_mat_dis}
 \frac{\text{d}}{\text{d}t}{\mathbf{U}}_{ijk}+\mathcal{D}_{ijk}\left({\mathbf{U}}\right) = 0,
\end{equation}
where ${\mathbf{U}}_{ijk}$ represents the vector of conservative variables stored at cell centres,  and 
\begin{eqnarray}
 \mathcal{D}_{ijk} :=&& \frac{1}{\Delta x^x_{ijk}}\left(\widetilde{\mathbf{F}}^x_{i+1/2,jk}-\widetilde{\mathbf{F}}^x_{i-1/2,jk}\right)\nonumber\\
&+&\frac{1}{\Delta x^y_{ijk}}\left(\widetilde{\mathbf{F}}^y_{i,j+1/2,k}-\widetilde{\mathbf{F}}^y_{i,j-1/2,k}\right)\nonumber\\
&+&\frac{1}{\Delta x^z_{ijk}}\left(\widetilde{\mathbf{F}}^z_{ij,k+1/2}-\widetilde{\mathbf{F}}^z_{ij,k-1/2}\right) - \mathbf{s}_{\text{\scriptsize{non-con.}},ijk},\label{eq:spat_op}
\end{eqnarray}
where $\widetilde{\mathbf{F}}^d_{m\pm1/2}$, for $m=i,j,k$, are the cell-wall numerical flux functions in the direction $d = x,y,z$. 
The numerical fluxes are computed through successive sweeps of each spatial dimension and summed according to equation \eqref{eq:spat_op}, resulting in a spatially unsplit method. 

The fluxes for the thermodynamic variables are computed using the HLLC solver developed by \citet{ToroHLLC} (note that the full multiphase system uses the HLLD solver from \citet{Barton2019}). To achieve higher order spatial accuracy, the initial conditions for the Riemann solver are taken to be the piecewise linear (MUSCL) reconstruction of the cell centred primitive variables. Primitive reconstruction is employed here so as to be consistent with the full multiphase system, but conservative reconstruction is also possible for the reduced system. An interface reconstruction method is also applied to reduce numerical diffusion around interfaces. This is achieved using the Tangent of Hyperbola INterface Capturing (THINC) method: an algebraic interface reconstruction technique that fits a hyperbolic tangent function to variables inside a cell. Specifically, this work uses the MUSCL-BVD-THINC scheme of \citet{BVDTHINC}. This scheme provides an additional check to minimise oscillations by comparing the reconstructed state's cell boundary variation with the previously calculated MUSCL reconstruction. THINC-reconstructed states are only accepted when their total boundary variation is lower than that of the MUSCL scheme alone. This procedure is applied to all variables, including the rigid volume fraction field. 

The fluxes for the rigid body volume fraction are calculated using a simple upwind method, based on the sign of $\vb{v}_r$ and the cell-wall reconstructed values of $r$: $r_L$ and $r_R$. For example, in the $x$-direction, let $v_r$ be the $x$-component of the rigid body velocity at the cell-wall, $(\mathbf{v}_r)_{i+1/2,jk}$, then the flux for the rigid volume fraction is given by:
\begin{align}
\widetilde{\mathbf{F}}^x_{i+1/2,jk}(r) = \left\lbrace\mqty{r_{L}v_r && \text{ if } v_r > 0 \\
							   r_{R}v_r && \text{ if } v_r < 0 \\}\right.
\end{align}

If greater accuracy is required for the advection of the rigid body field, this method can be replaced by any desired scheme, such as volume of fluid methods. The rigid body volume fraction is not a thermodynamic variable, so it may be updated completely separately to the other variables. Thermodynamic variables do not need to be updated in rigid ($r > 0.5$) cells, saving computational expense, as they will be replaced by the interface seeding method in Section \ref{sec:RigidBodyMethod}.

Finally, the non-conservative term for the rigid body volume fraction is calculated. This term is added using a simple difference formula, similar to previous works \cite{Barton2019} for consistency. This term can also be written as a sum of `fluxes'. In what follows, variables with subscript $L$ or $R$ are evaluated at the cell wall on either side of a cell in a given direction. For example, in the x direction $L \rightarrow (i-1/2,jk)$ and $R \rightarrow (i+1/2,jk)$. All other variables are evaluated at the cell centre $(ijk)$. The non-conservative term can then be written:
\begin{eqnarray}
 \mathbf{s}_{\text{\scriptsize{non-con.}}} :=&& \frac{1}{\Delta x^x}\left(\widetilde{{\cal F}}^x_{R}-\widetilde{{\cal F}}^x_{L}\right)\nonumber\\
&+&\frac{1}{\Delta x^y}\left(\widetilde{{\cal F}}^y_{R}-\widetilde{{\cal F}}^y_{L}\right)\nonumber\\
&+&\frac{1}{\Delta x^z}\left(\widetilde{{\cal F}}^z_{R}-\widetilde{{\cal F}}^z_{L}\right) \ ,
\end{eqnarray}
where $\widetilde{{\cal F}}^x_{R}(r) = r\times(\vb{v}_r)^x_R$ and $\widetilde{{\cal F}}^x_L = r\times(\vb{v}_r)^x_L$, with similar definitions for the other directions.

When the full multiphase system is used, the same form for the non-conservative update is used for the material volume fraction and stretch tensor update. For the material volume fraction, in direction $d$:
\begin{align}
\widetilde{{\cal F}}^d_{L}(\phi_{(l)}) = \phi_{(l)}\times(u^*)^d_L
\end{align}
where $u^*$ is the velocity at the cell-wall, found from the HLLD Riemann solver in the $d$ direction. For the stretch tensor:
\begin{align}
 \widetilde{{\cal F}}^d_{L}(\Vbar_{ab}) =  \frac{2}{3}\Vbar_{ab}\times(u^*)^d_L  - u_a\times\left({\Vbar_{db}}^*\right)^d_L \ ,
\end{align}
where ${\Vbar}^*$ is the stretch tensor at the cell-wall, found from the HLLD Riemann solver in the $d$ direction.

To achieve a higher temporal resolution in the update of the hyperbolic terms, the third order Runge-Kutta time integration scheme is used:
\begin{eqnarray}
\mathbf{U}^{(1)}    &=& \mathbf{U}^{n} - \Delta t  \mathcal{D}\left({\mathbf{U}}^{n}\right)\\
\mathbf{U}^{(2)}    &=& \mathbf{U}^{(1)} - \Delta t  \mathcal{D}\left({\mathbf{U}}^{(1)}\right)\\
\mathbf{U}^{(3)} &=& \frac{3}{4} \mathbf{U}^{n} + \frac{1}{4} \mathbf{U}^{(2)} \\
\mathbf{U}^{(4)}    &=& \mathbf{U}^{(3)} - \Delta t  \mathcal{D}\left({\mathbf{U}}^{(3)}\right)\\
\mathbf{U}^{n+1} &=& \frac{1}{3} \mathbf{U}^{n} + \frac{2}{3} \mathbf{U}^{(4)} \ .
\end{eqnarray}

\subsection{Rigid body method}
\label{sec:RigidBodyMethod}
Rigid bodies introduce undeforming, perfectly reflective boundaries within the domain, also known as immersed boundaries, which are governed by the evolution equation \eqref{eq:revo}.
The boundary condition at rigid body interfaces has two components: no-penetration and either slip or no-slip. At the interface, the no-penetration condition requires:
\begin{equation}
(\mathbf{u}-\mathbf{v}_r)\cdot\normal_r = 0 \ ,
\end{equation}
where $\normal_r$ is the rigid body normal vector. In other words, the velocity aligned with the rigid interface normal component must equal the rigid body velocity in that direction. For the slip boundary condition, all other tangential velocity components remain unchanged across the interface. The no-slip condition is not employed here, but is enforced by the entire velocity vector to equal the rigid body velocity at the interface, resulting in the condition $\mathbf{u} = \mathbf{v}_r$. Therefore, a method is required that will fill the rigid cells ($r > 0.5$) with a suitable thermodynamic state such as to reproduce these boundary conditions.

The method presented here shall be applicable to both the reduced system and the full multi-material system. Analogously to the method in \citet{WallisFluxEnriched}, the rigid body boundary conditions are mediated by flux-modifiers and interface seeding routines. However, by comparison, no new flux modifiers are required: only an interface seeding method is required for rigid body interaction. This is because the rigid body boundary only involves one real material, with the flux inside the rigid body not being relevant, therefore meaning that a two-flux method such as those presented by \citet{WallisFluxEnriched} is not required.

Although no additional flux-modifiers are required, rigid bodies must be seeded with suitable states to enable the correct boundary interaction, as previously mentioned. This is done using an interface seeding routine, similar to those presented in \citet{WallisFluxEnriched}. This procedure is carried out at the start of every time-step, before the calculation of the hyperbolic update.

In the following, it is necessary to distinguish between variables that will be reflected across a rigid body interface and those that will not, denoted by $\mathbf{U}^{\reflected}$ and $\mathbf{U}^{\constant}$ respectively. All scalar quantities are constant across the rigid interface. For vector and tensor quantities, it is easiest to define these quantities in the reference frame where $\normal_r$ points in the $x$-direction $\mathbf{\hat{x}}$ (although analogous results can be obtained for any direction). States that have been rotated into this reference frame will be denoted with a tilde, $\mathbf{U} \rightarrow \tilde{\mathbf{U}}$, and are found in the normal way, where for a vector $\mathbf{v}$ and matrix $\mbox{M}$:
\begin{align}
 \mathbf{\tilde{v}} &= \mathbf{R}\mathbf{v} \\
 \tilde{\mbox{M}} &= \mathbf{R}\mbox{M}\mathbf{R}^{\text{T}} \ .
\end{align}
The rotation matrix $\mathbf{R}$ is detailed in \citet{WallisFluxEnriched}.

Having rotated a state into this new frame, the reflected and constant parts of the velocity and stretch tensor are defined as:
\begin{align}
 \mathbf{\tilde{u}}^{\reflected}&=\mqty(\tilde{u}_x \\ 0 \\ 0 ) , \quad \tilde{\mathbf{V}}^{\reflected} = \mqty(0 && \tilde{V}_{xy} && \tilde{V}_{xz} \\ \tilde{V}_{yx} && 0 && 0 \\ \tilde{V}_{zx} && 0 && 0) \\
 \mathbf{\tilde{u}}^{\constant}	&=\mqty(0 \\ \tilde{u}_y \\ \tilde{u}_z ) , \quad \tilde{\mathbf{V}}^{\constant} = \mqty(\tilde{V}_{xx} && 0 && 0 \\ 0 && \tilde{V}_{yy} && \tilde{V}_{yz} \\ 0 && \tilde{V}_{zy} && \tilde{V}_{zz}) \ .
\end{align}

The interface seeding routine proceeds as follows:\\

\framebox[\textwidth]{\begin{minipage}{0.9\textwidth}{
The rigid seeding routine:
\begin{enumerate}
 \item In all rigid cells (with $r > 0.5$), the outward pointing interface normal, $\normal_r$, is calculated.
 \item A probe is sent out along the normal direction. The probe can either simply be taken to be a constant length, such as 1.5$\dd x$, or sample points can chosen at various distances stepping out in the normal direction (from $\dd x$ to 4$\dd x$), and when the $2^{\text{DIM}}$ enclosing cell centres around the sample point are valid ($r < 0.5$), this point is chosen. Once the probe point is chosen, material values are interpolated at that point to give a new state $\mathbf{U}_{\text{Interp}}$. If a valid probe position cannot be found, no further action is taken.
 \item  The interpolated cell is rotated so the normal lies in the $\mathbf{\hat{x}}$ direction, giving $\tilde{\mathbf{U}}_{\text{Interp}}$.
 \item A new state, $\tilde{\mathbf{U}}^{\mbox{\scriptsize{new}}}$, is then created by reflecting the suitable components of the interpolated cell:
\begin{equation*}
  \tilde{\mathbf{U}}^{\mbox{\scriptsize{new}}}  = \tilde{\mathbf{U}}_{\text{Interp}}^{\constant} - \tilde{\mathbf{U}}_{\text{Interp}}^{\reflected}\ .
 \end{equation*}
 If the rigid body is dynamic, then the normal component of the rigid body velocity must also be included. This is done adding $ 2 \mathbf{\tilde{v}}_r^{\reflected} = 2 (\mathbf{v}_r \cdot \normal_r)$ to the reflected velocity of $\tilde{\mathbf{U}}^{\mbox{\scriptsize{new}}}$. Therefore, in total:
 \begin{equation*}
  \mathbf{\tilde{u}}^{\mbox{\scriptsize{new}},\reflected}  = 2 \mathbf{\tilde{v}}_r^{\reflected} - \mathbf{\tilde{u}}_{\text{Interp}}^{\reflected}  \ .
 \end{equation*}
 \item Here, the method then diverges with two options: the original or Riemann problem based version.
 \begin{itemize}
  \item For the original method, this new state is then rotated back into the original frame of reference and replaces the original state.
 \begin{equation*}
  \mathbf{U} \leftarrow \mathbf{U}^{\mbox{\scriptsize{new}}} \ .
 \end{equation*}
 \item For the Riemann problem based method, the HLLC(D) star states are calculated using the same method as for the flux calculation, with the left and right states being $\tilde{\mathbf{U}}^{\mbox{\scriptsize{new}}}$ and $\tilde{\mathbf{U}}_{\text{Interp}}$ respectively. This gives the intermediate states $\tilde{\mathbf{U}}_L^*$ and $\tilde{\mathbf{U}}_R^*$. The $\tilde{\mathbf{U}}_R^*$ state is then rotated back into the original frame of reference and replaces the original state.
  \begin{equation*}
  \mathbf{U} \leftarrow \mathbf{U}_R^* \ .
 \end{equation*}
 \end{itemize}
\end{enumerate}
}\end{minipage}}\\[0.5cm]

This procedure will then fill a few layers of cells inside the rigid body interface with a state that will produce the desired rigid body boundary condition, and the hyperbolic update can proceed as normal.

Although it has been mentioned that no new flux-modifiers are required for the method, the existing void-opening flux modifier can be applied to rigid-solid interfaces to allow solid bodies to collide with and separate from rigid bodies, just as they would with any other solid body in the domain. This then allows for void opening in solid-rigid collisions. The flux modifier is applied when the rigid volume fraction `changes sign' between the two cells being used to calculate the flux:
\begin{equation}
 (0.5-r_L)(0.5-r_R) < 0 \ .
\end{equation}

\subsection{Normal Estimation}

Two different methods are considered for the calculation of the rigid body normal from the volume fraction field $r$. The first is from \citet{YoungsNormal}. Youngs' method is a first order approximation to the normal, attained using a simple difference formula for the local $3^{\text{\scriptsize{DIM}}}$ cells. The method is cheap, and is easily extended into three dimensions. Following the notation of \citet{YoungsNormal}, the normal of a field $f$ is estimated as:
\begin{align}
 \hat{n} = - \frac{\grad f}{|\grad f|} \ ,
\end{align}
with
\begin{align}
\pdv{f}{x} &= \frac{f_E - f_W}{2\dd x} \\
\pdv{f}{y} &= \frac{f_N - f_S}{2\dd y} \\
\pdv{f}{z} &= \frac{f_T - f_B}{2\dd z} \ ,
\end{align}
where symmetric difference formulae must be used for each direction. These formulae are outlined by \citet{YoungsNormal} and \citet{ELVIRA}.

Although Youngs' method is efficient, it is only first order, and it is not guaranteed to reproduce linear interfaces exactly.

The second order ELVIRA method is also considered. This method is outlined by \citet{ELVIRA}, and produces a second order accurate interface normal. The method attempts to find the best normal from a set of candidate normals that are again obtained using simple difference formulae. The measure of a candidate normal's quality found by calculating the volume fractions the candidate normal would produce in neighbouring cells, and comparing these volume fractions to the true cell volume fractions. Whichever candidate normal has the closest match is then selected. This results in a more accurate calculation of the normal, with linear interfaces being reconstructed exactly in cells where the interface passes through the centre cell \cite{ELVIRA}. Naturally, the ELVIRA method is more expensive as it requires the calculation of multiple candidate normals and calculation of geometric informations from the interface. As such, ELVIRA is only employed in two dimensional validation tests, with Youngs' method being used in three dimensions.

\section{Validation and Verification}

Throughout all the tests, the rigid body is plotted in solid grey for all rigid ($r>0.5$) cells.

\subsection{Shock Reflection From a Wedge}

Initially, two shock-wedge interaction tests are considered. These tests feature a shock wave in an ideal gas colliding with an angled plane. The first test features a Mach 1.7 shock colliding with a 25$^{\circ}$ wedge, producing a single reflection. The second test features a Mach 10 shock colliding with a 30$^{\circ}$ wedge, producing a more complex double reflection. These tests examine the ability of the model to give accurate solutions across different shock strengths. The Mach 1.7 test can be compared to the similar test in \citet{DumbserRigid}, and the Mach 10 test can be compared to the similar tests in \citet{DoubleMachReflection} and \citet{ForrerRigidBoundaries}.

\begin{figure}
\centering
\includegraphics[width = \textwidth]{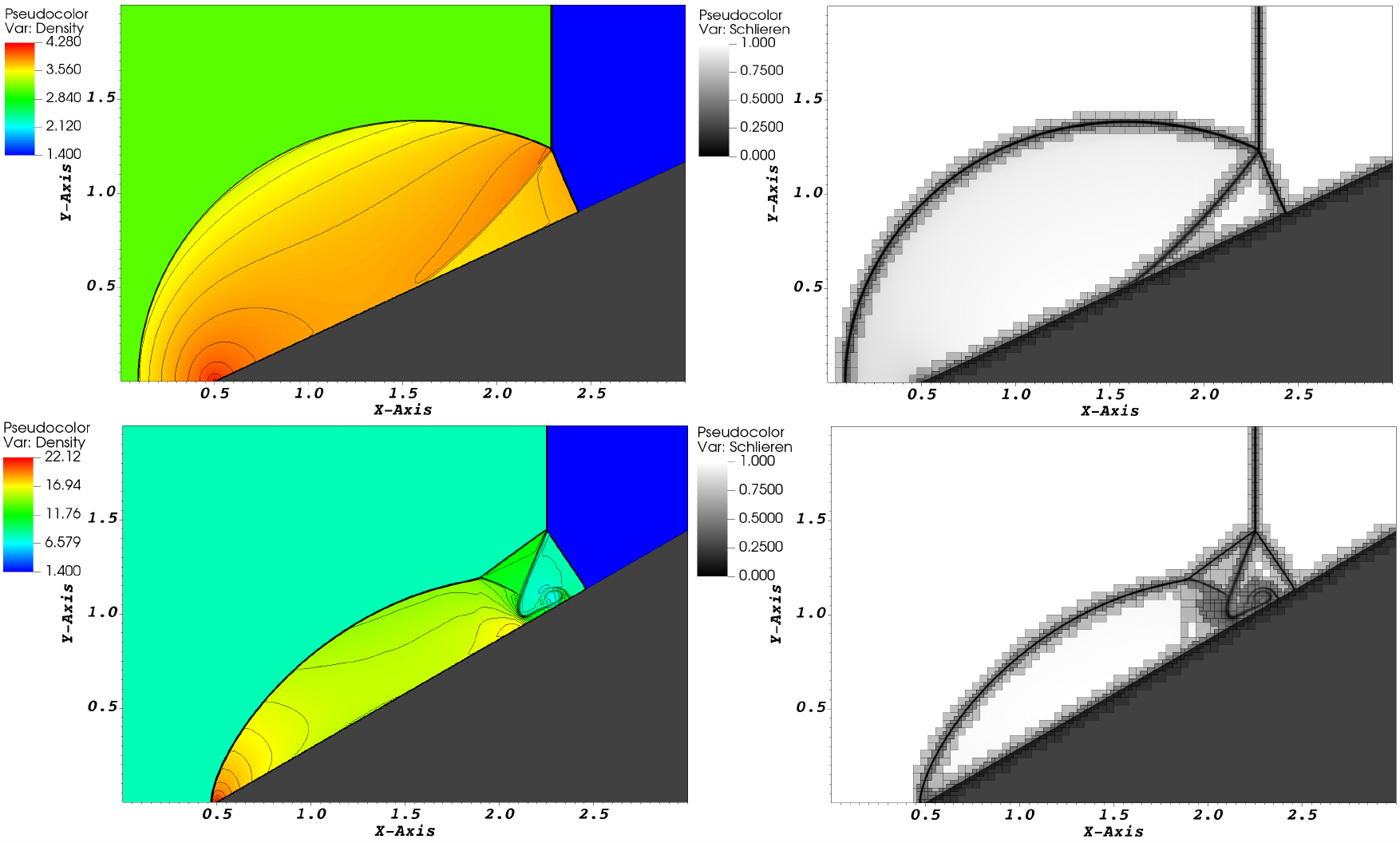}
\caption{Different shock reflections problems. The images show (\textit{left}) the density and (\textit{right}) a numerical schlieren with AMR levels. (\textit{top}) The Mach 1.7 single shock reflection test at $t = 1.2$ and (\textit{bottom}) the Mach 10 double shock reflection test at $t=0.2$. The rigid body is depicted in grey.}
\label{fig:Wedge}
\end{figure}

Both of these tests are non-dimensionalised, with a domain $x = [0:3]$, $y = [0:2]$, a base resolution of 300 $\times$ 200, and two levels of AMR, both refinement factor 2. The gas is a $\gamma = 1.4$ ideal gas, with a quiescent state of $\rho = 1.4, p = 1.0, \vb{u} = \vb{0}$. The rigid body wedge starts at $x = 0.5$ and the shock is initially located at $x = 0.25$ in both cases. A CFL of 0.4 is used, with the first test being run to a time of 1.2 and the second test being run to a time of 0.2. These tests are shown in Figure \ref{fig:Wedge}. The method handles the test well, replicating the results of previous numerical simulations. In this case, the second order ELVIRA normal estimation method is preferred over the first order Young's method. The ELVIRA method produces a much more accurate representation of the normal, which is constant from cell to cell, whereas the first order method varies slightly, causing oscillations near the interface.

\subsection{Shardin Wedge}
Similar to the previous test is the Shardin wedge problem, for which there is a range of experimental and numerical comparison. The original study by \citet{Schardin} took high-frequency schlieren photographs of a shock hitting a triangular wedge. This work examines the near identical case from \citet{ChangWedge}, in which a Mach 1.34 shock hits an equilateral triangle. The quantitative experimental comparison for the test is obtained by tracking the positions of different flow features over the course of the simulation. The positions are non-dimensionalised with respect to the triangle length and width to enable comparison to a broader range of tests. Similar experiments can be found in the literature: \cite{GhostPointForcingWedge, ChaudhuriWedge, DumbserRigid, UdaykumarRigid, SivierWedge}. 

\begin{figure}
\centering
\includegraphics[width = 0.8\textwidth]{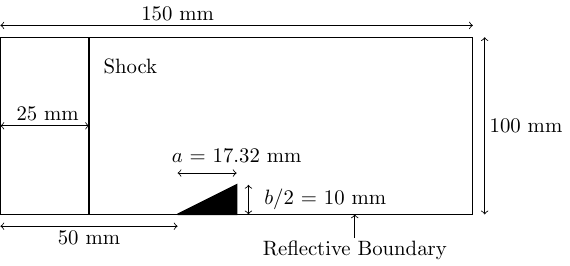}
\caption{The initial conditions for the Schardin wedge problem.}
\label{fig:WedgeInitialConditions}
\end{figure}

The initial set-up for this test is shown in Figure \ref{fig:WedgeInitialConditions}. The test is run for 0.35 ms, with a CFL of 0.4. A base resolution of $150 \times 100$ was used, along with 4 layers of AMR, each of refinement factor 2. The gas is again a $\gamma=1.4$ ideal gas, with the quiescent state given by: $\rho = 1 \ \text{kgm}^{-3}, p = 0.05 \ \text{MPa}, \vb{u}=\vb{0}$.

\begin{figure}
\centering
\includegraphics[width = 0.9\textwidth]{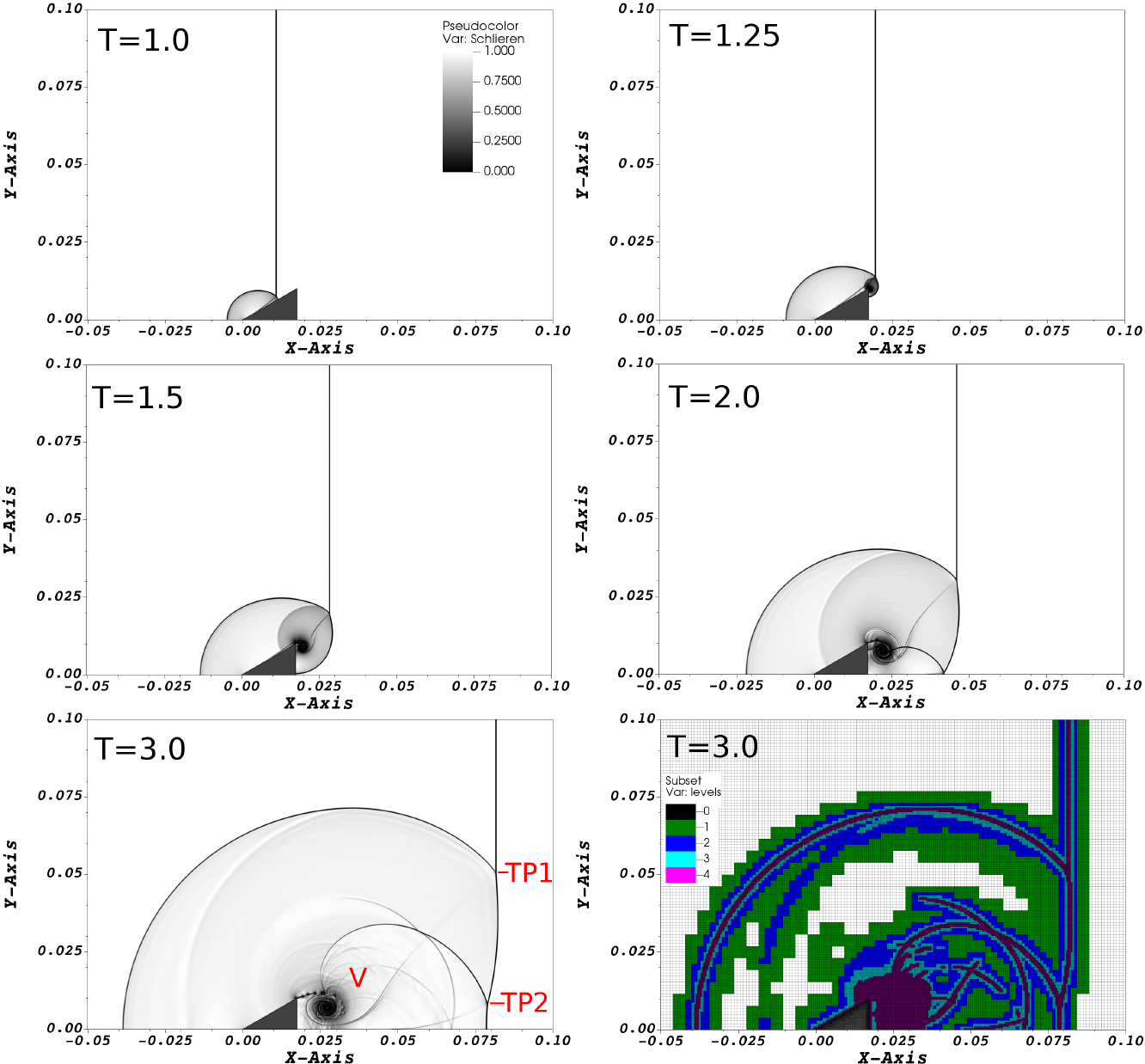}
\caption{The Schardin wedge test. The images show a numerical schlieren of the solution, with times given in $1\times 10^{-4}$s. The last image shows the AMR levels used. Experimental comparison is drawn by tracking the position of different flow features: the upper shock triple point (TP1), the lower shock triple point (TP2) and the vortex core (V).}
\label{fig:SchardinWedgeImages}

\includegraphics[width = 0.75\textwidth]{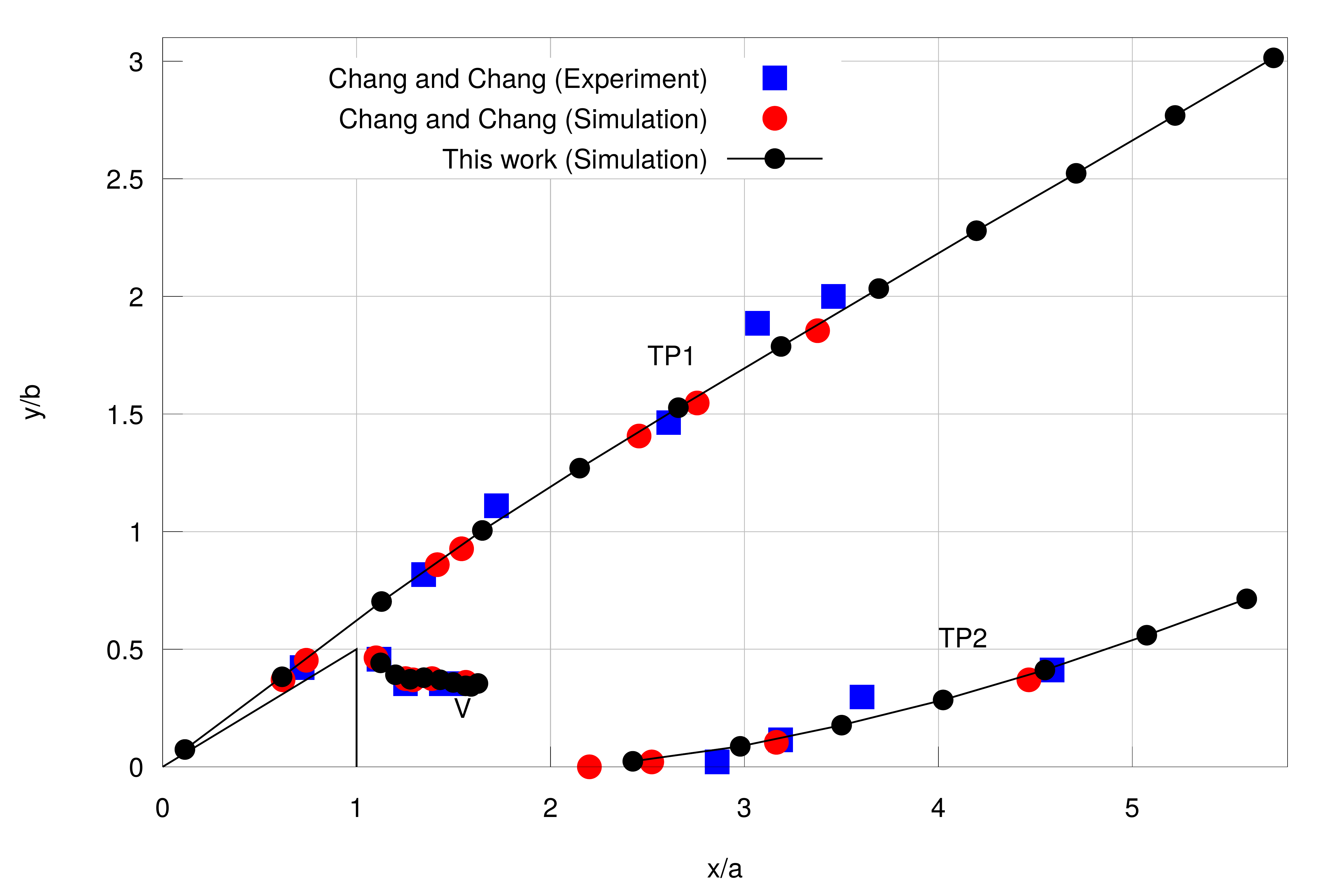}
\caption{Comparison of the present work with the previous experiments and simulations of \citet{ChangWedge}. The locations of the upper triple point (TP1), lower triple point (TP2) and the vortex core are tracked over the course of the simulation. Excellent agreement is observed. Each axis is non-dimensionalised with respect to the dimensions of the triangle, $a$ and $b$, outlined in Figure \ref{fig:WedgeInitialConditions}.}
\label{fig:SchardinWedgeExperiment}
\end{figure}

The results of the test and the experimental comparison are shown in Figure \ref{fig:SchardinWedgeImages} and Figure \ref{fig:SchardinWedgeExperiment}. Excellent agreement with the results of \citet{ChangWedge} is observed.

\subsection{Cylinder Test}

Another shock-body interaction problem is that of a cylinder being hit by a shock wave. This test has been studied by many other authors \cite{Bryson, ChaudhuriWedge, Kaca, UdaykumarRigid, Zoltak}, with both numerical and experimental comparison, making it a suitable validation case.

The test is non-dimensionalised, with a domain spanning $x=[0:2]$, $y=[0:1]$ and a reflective boundary condition along both the upper and lower $x$-axis. The cylinder has a radius of 0.105 and is centred at $(0.5,0)$. The test is run until $t=0.4$ with a CFL of 0.4. A base resolution of 200 $\times$ 100 is used, with 4 levels of AMR, all refinement factor 2. A Mach 2.81 shock is initialised at $x=0.385$, with the quiescent state being a $\gamma = 1.4$ ideal gas with $\rho = 1, p=1, \vb{u} = \vb{0}$.

This test suffers from the carbuncle instability if the standard HLLC approximate Riemann solver is employed. To avoid this issue, the small HLLC-LM modification from \citet{HLLCLM} is used in this test.

\begin{figure}
\centering
\includegraphics[width = \textwidth]{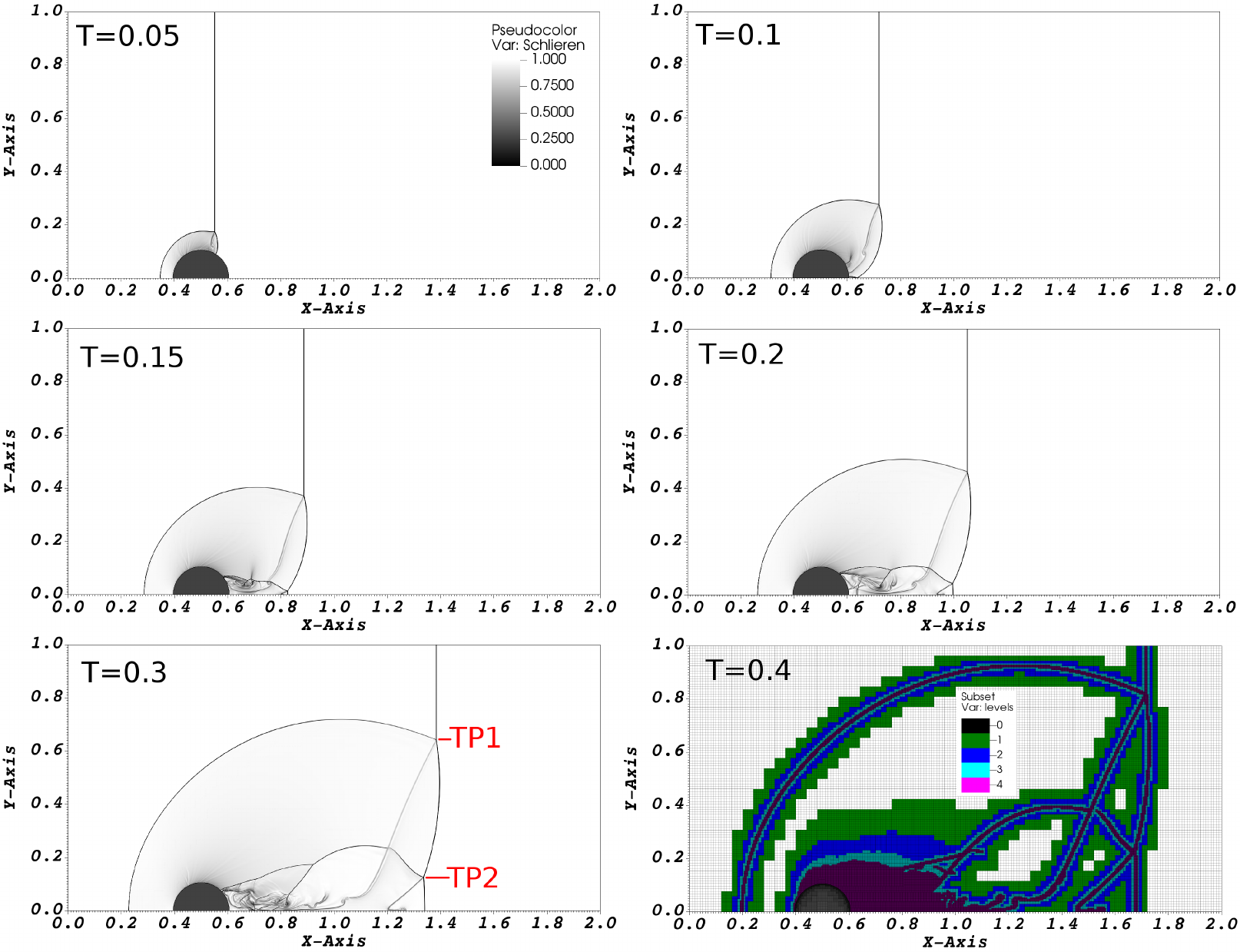}
\caption{The cylinder test. The images show a numerical schlieren of the solution, with the last image showing the AMR levels used. Experimental and numerical comparison is drawn by tracking the position of different flow features: the upper shock triple point (TP1) and the lower shock triple point (TP2).}
\label{fig:Cylinder}
\includegraphics[width = 0.8\textwidth]{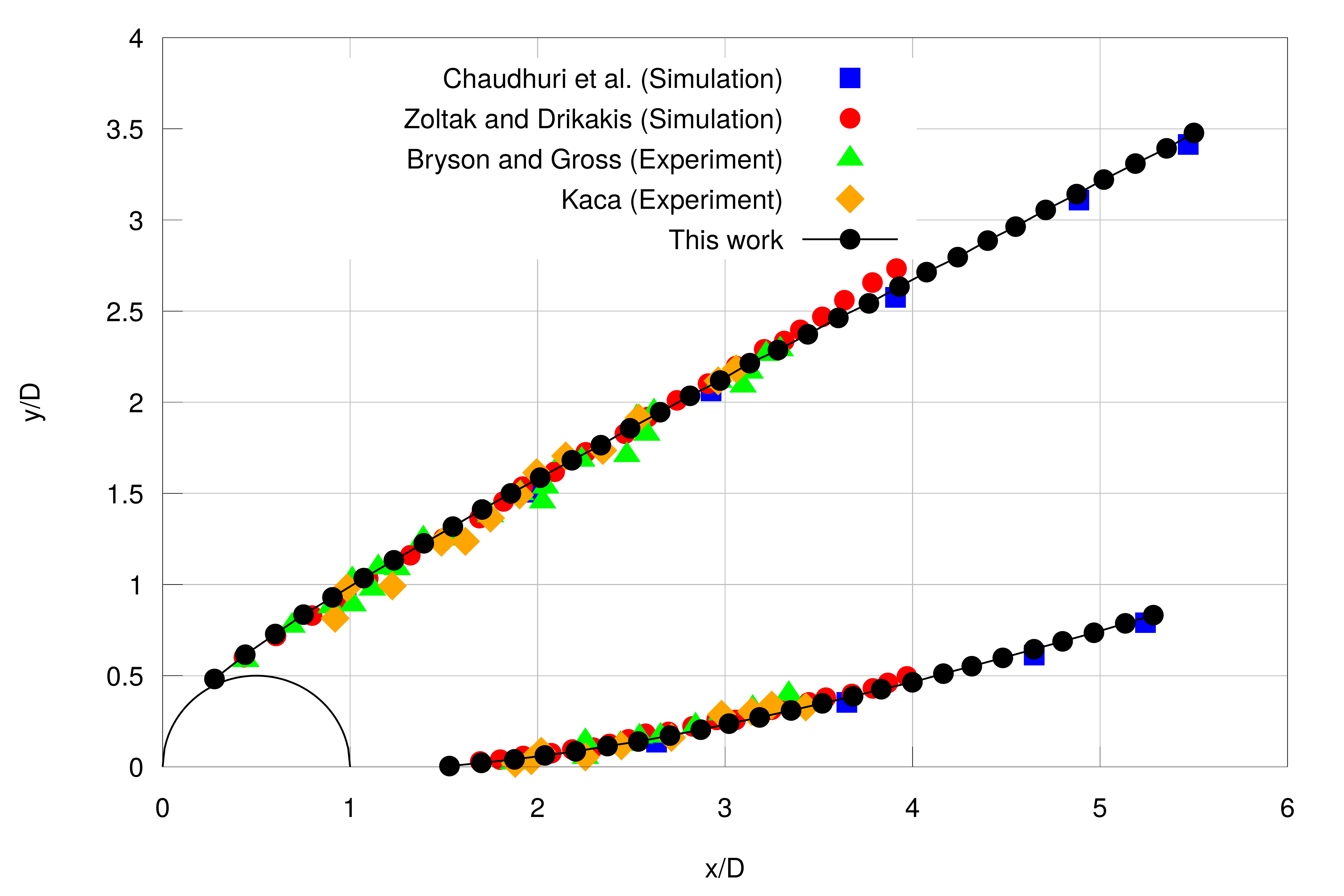}
\caption{Comparison of cylinder test with the previous experiments and simulations. The locations of the upper triple point (TP1) and lower triple point (TP2) are tracked over the course of the simulation. Excellent agreement is observed with both the previous simulations of \citet{ChaudhuriWedge} and \citet{Zoltak}, and the experimental studies of \citet{Bryson} and \citet{Kaca}. Each axis is non-dimensionalised with respect to the diameter of the cylinder, $D$.}
\label{fig:CylinderComparison}
\end{figure}

The test is shown in Figure \ref{fig:Cylinder}, where the images depict a numerical schlieren of the density profile at different times. Figure \ref{fig:CylinderComparison} shows the numerical and experimental comparison with previous works, and excellent agreement is observed.

\subsection{Moving Cylinder Test}

To test the ability of the model to handle dynamic rigid bodies, a moving cylinder is considered. The test is again non-dimensionalised, with the surrounding gas being a $\gamma = 1.4$ ideal gas with $\rho=1.4, p=1, \vb{u}=\vb{0}$. The domain for this test is $x=[0:5], \ y=[-2:2]$, with a resolution of 400 $\times$ 320. The rigid cylinder has a radius of 0.5 and is initially located at $(x,y) = (0.6,0)$. The velocity of the rigid body is constant at $\mathbf{v}_r = 3 \ \hat{x}$. The test is run until $t=1$ with a CFL of 0.4.

\begin{figure}
\centering
\includegraphics[width = \textwidth]{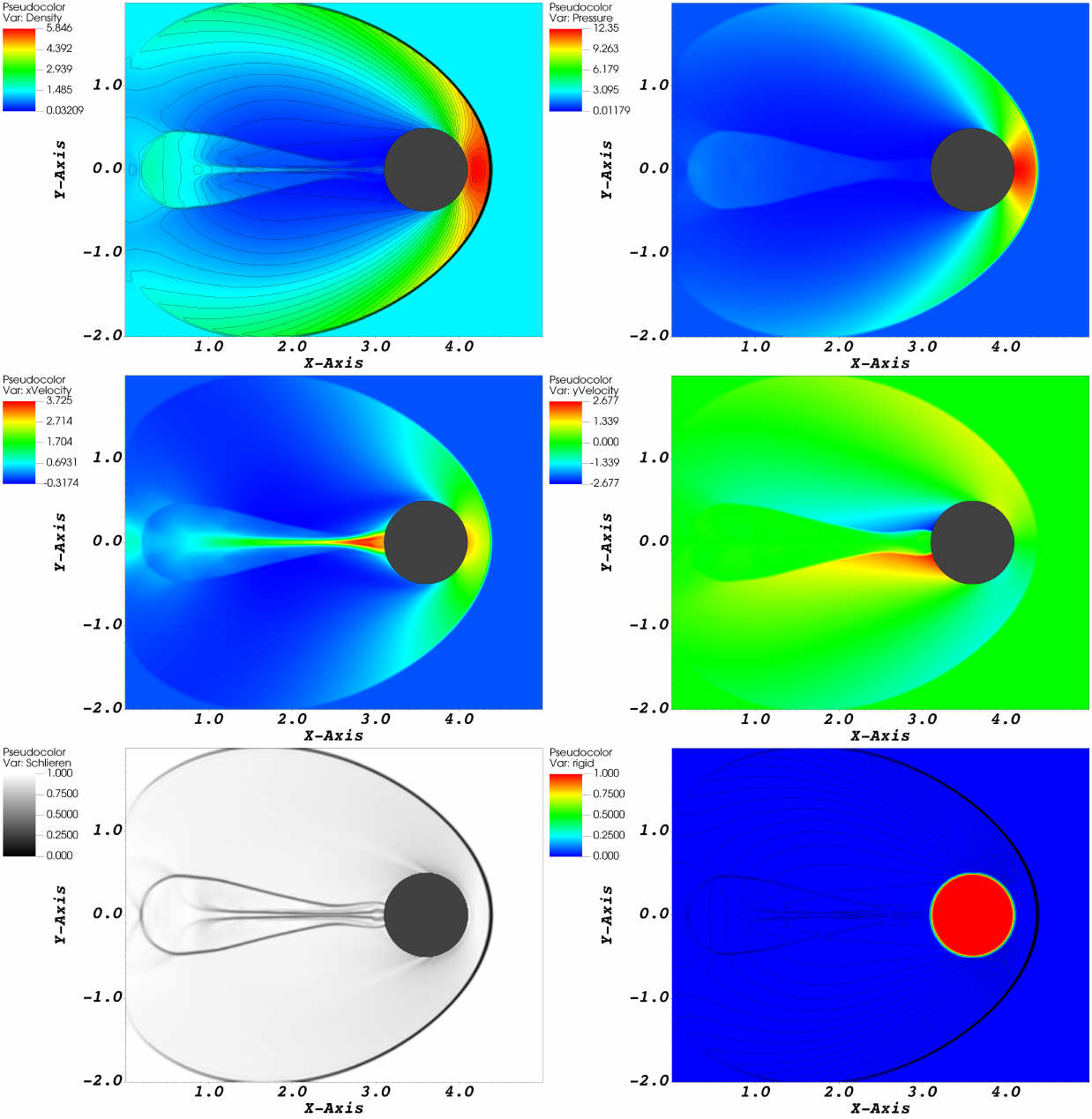}
\caption{The moving cylinder test. The images show (from left to right and top to bottom) density, pressure, $x$-velocity, $y$-velocity, a numerical schlieren and the rigid volume fraction.}
\label{fig:MovingCylinder}
\end{figure}

The test is shown in Figure \ref{fig:MovingCylinder}, depicting density, pressure, $x$-velocity, $y$-velocity, a numerical schlieren and the rigid volume fraction. The rigid body volume fraction is kept sharp thanks to the THINC reconstruction performed in its update, and the test demonstrates the ability of the method to handle supersonic dynamic rigid body motion.

\begin{table}
\begin{center}
    \begin{tabular}[t]{|l|l|l|l|l|l|l|}
      \hline
      Material & $\rho_0$ \ kgm$^{-3}$ & $K_0$ \ GPa & $G_0$ \ GPa & $\alpha$ & $\beta$ & $\Gamma_0$ \\
      \hline
      Aluminium     & 2703.0 & 76.3  & 26.36& 0.627& 2.288& 1.484 \\
      Copper        & 8960.0 & 130.1 & 43.33& 1.0  & 3.0  & 2.0   \\
      Steel         & 8030.0 & 156.2 & 77.2 & 0.569& 2.437& 1.563 \\
      \hline
    \end{tabular}
\end{center}
\caption{Equation of state material parameters for equation \eqref{eq:Romenskii}.}
\label{tab:Romenskii}
\end{table}

\begin{table}
\begin{center}
    \begin{tabular}[t]{|l|l|l|l|l|l|l|l|}
      \hline
      Material & $c_1$  GPa & $c_2$  GPa & $c_3$ & $n$ & $m$ & $T_{\text{melt}}$  K & $C_V$ J kg$^{-1}$K$^{-1}$  \\ \hline
      Copper   & 0.09 & 0.292 & 0.025 & 0.31  & 1.09 & 1358.0 & 385.0  \\
      \hline
    \end{tabular}
\end{center}
\caption{Johnson Cook plasticity parameters.}
\label{tab:PlasticParameters}
\end{table}

\subsection{Solid Extrusion with a Piston}

This test, proposed in \citet{HillSolidWedge}, examines the interaction of both static and dynamic rigid bodies with an elastoplastic solid. The test consists of a rigid piston pushing a block of aluminium through a rigid wedge, representing an extrusion problem. The presence of the elastoplastic solid in this case requires the use of the full multi-phase system of equations. The test is more complex than the fluid-wedge interaction problems, both because of the driving piston and the separation of the elastic precursor from the plastic wave. The aluminium in this test is given by the equation of state equation \eqref{eq:Romenskii} with parameters given in Table \ref{tab:Romenskii}. Following \citet{HillSolidWedge}, the aluminium has an ideal plasticity law with the yield stress given by $\sigma_Y = 0.2976$ GPa.

\begin{figure}
\centering
\includegraphics[width = \textwidth]{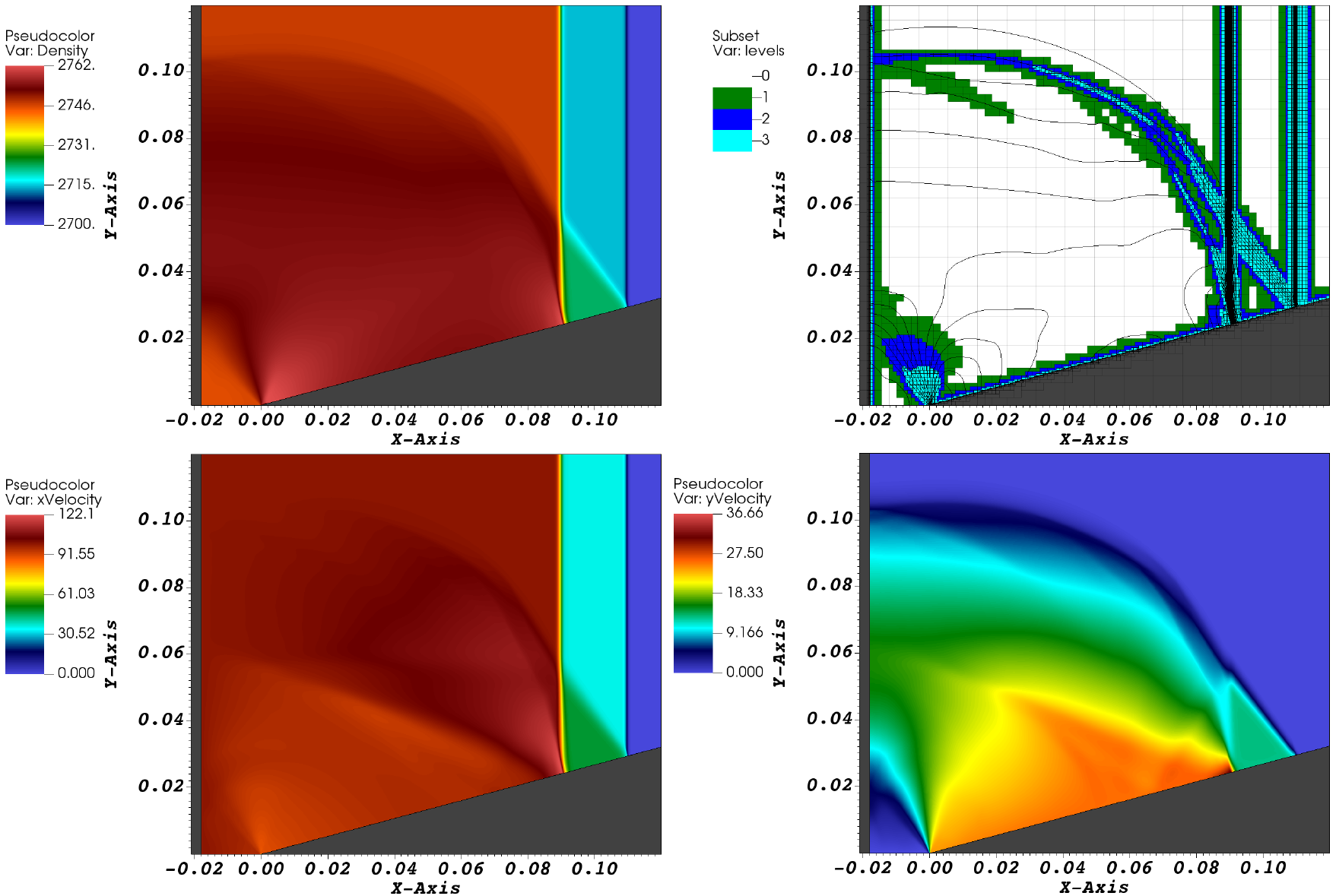}
\caption{The solid extrusion with a piston test at t = 20 $\mu$s. The images show (from top to bottom, left to right) the density, AMR grids and density contours, $x$-velocity and $y$-velocity. The test demonstrates the stability of the method when handling a combination of elastoplastic solids and both static and dynamic rigid boundaries.}
\label{fig:SolidWedge}
\end{figure}

The domain for this test is $x = [-21:120]$ mm, $y = [0:120]$ mm, with a reflective upper boundary. A base resolution of $256 \times 216$ is used, with 3 levels of AMR, refinement factor 2. The test is run for 20 $\mu$s with a CFL of 0.4. The piston is initially located at $x = -20$ mm and given a constant velocity of $\vb{v}_r = 100 \hat{x}$ ms$^{-1}$. The wedge has an angle of 15$^{\circ}$ with respect to the $x$-axis and starts at $x = 0$. The method copes well with this strenuous problem, as can be seen from the results in Figure \ref{fig:SolidWedge}.

\subsection{Solid Impact Extrusion}

Another example of the use of a rigid body and elastoplastic solid formulation together is the impact extrusion test from \citet{GrayImpactExtrusion}. In this test, a copper sphere impacts an extrusion cone, resulting in a jet of metal being ejected. The jet has a much higher velocity than the sphere, and undergoes necking, breaking into fragments. The test can be compared with both the results of \citet{GrayImpactExtrusion} and other works in the literature \cite{BonoraImpactExtrusion, HornqvistImpactExtrusion, IannittiImpactExtrusion, ParkImpactExtrusion, UdaykumarGroove}.

This test is an example of the particular usefulness of the formulation at hand. In experiment, the hardened steel extrusion cone is not expected to deform significantly, meaning it can be replaced with a rigid body in numerical studies, greatly reducing the cost of the test. Additionally, the experiment is run surrounded by void using the model of \citet{WallisFluxEnriched}. This removes the need to track the high speed waves in any surrounding fluid which ultimately have little effect on the motion of the elastoplastic solid.

\begin{figure}
\centering
\begin{subfigure}{0.49\textwidth}
 \centering
 \includegraphics[width = 0.4\textwidth]{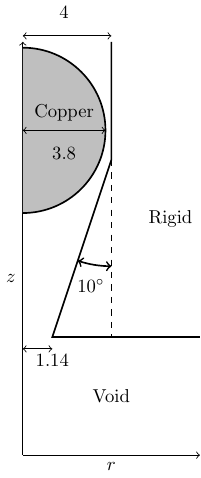}
 \caption{~}
 \label{fig:ImpactExtrusionInitialConditions}
\end{subfigure}\hspace{1mm}
\begin{subfigure}{0.49\textwidth}
 \centering
 \includegraphics[width = \textwidth]{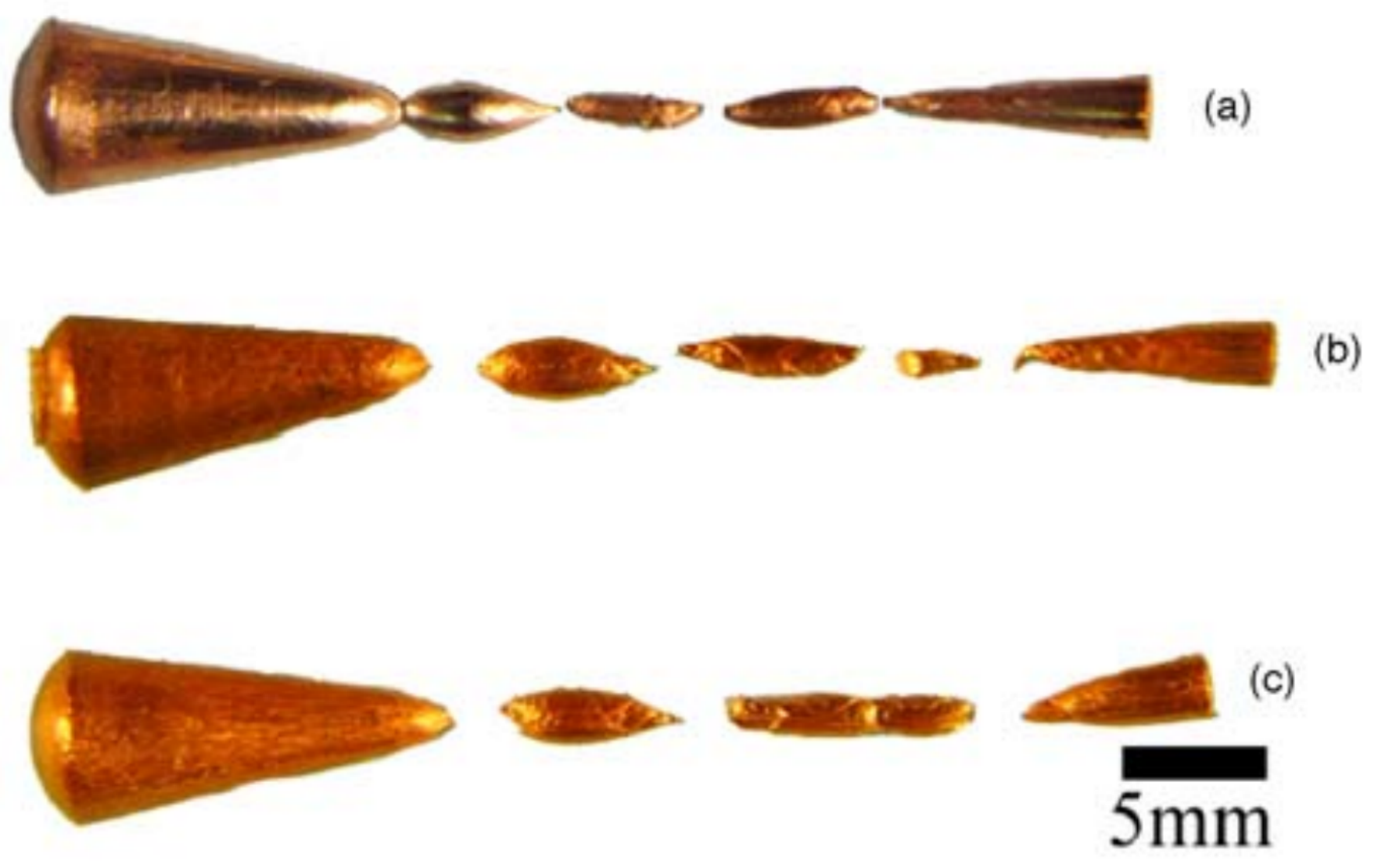}
 \caption{~}
 \label{fig:ImpactExtrusionExperiment}
\end{subfigure}
\caption{The impact extrusion test. (a) The elastoplastic impact extrusion test initial conditions. Lengths shown are in mm. In the domain, the base of the rigid body is positioned at $z=0$. (b) The experimental results of elastoplastic impact extrusion test performed by \citet{GrayImpactExtrusion}, presenting three different copper grain sizes. Reproduced from Gray et al., Influence of Shock Prestraining and Grain Size on the Dynamic-Tensile-Extrusion Response of Copper: Experiments and Simulation, \textit{AIP Conference Proceedings}, 845, 725-728 (2006) \cite{GrayImpactExtrusion}, with the permission of AIP Publishing.}
\label{fig:ImpactExtrusionCombined}
\end{figure}

\begin{figure}
\centering
\includegraphics[width = \textwidth]{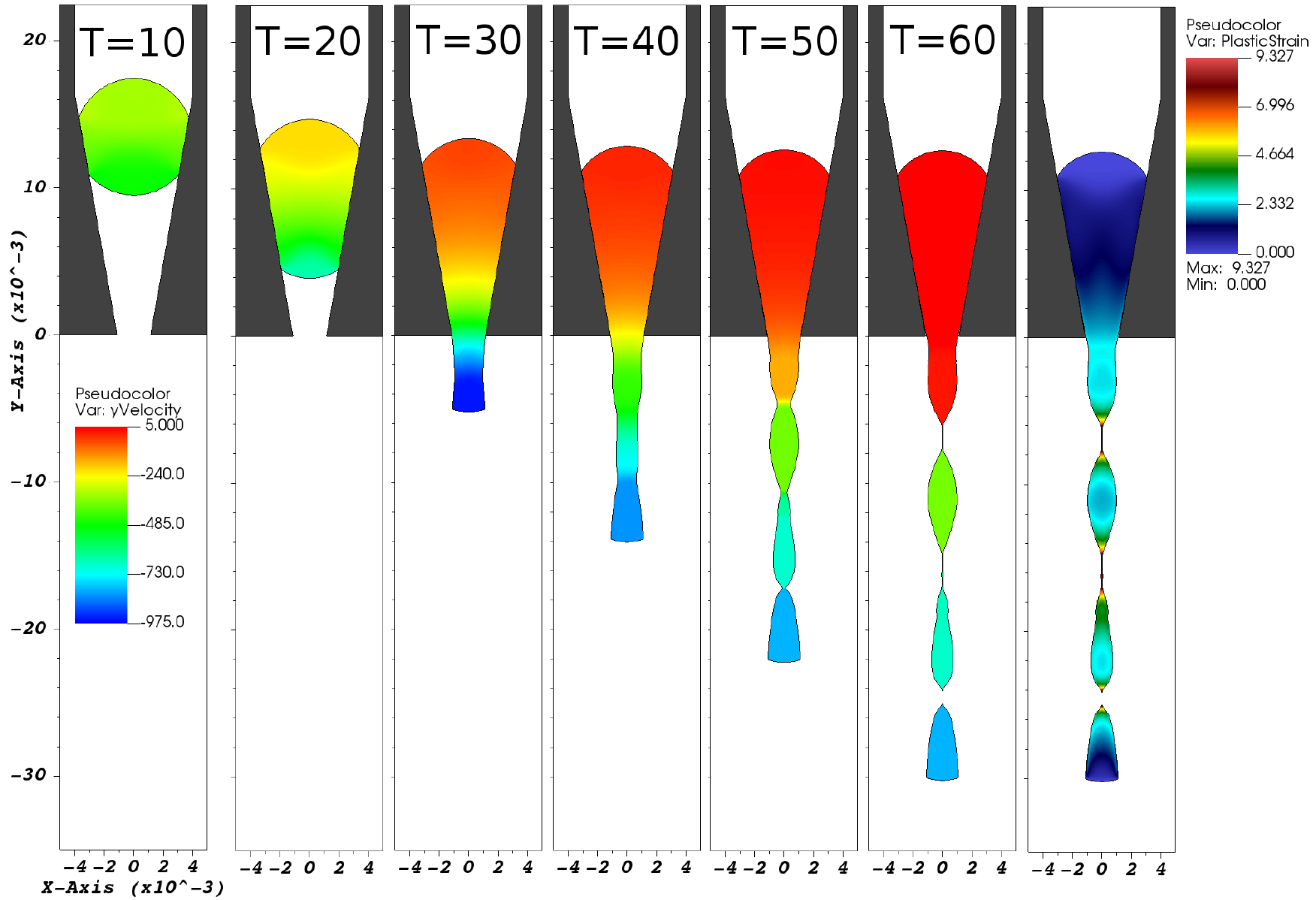}
\caption{The elastoplastic impact extrusion test. An elastoplastic sphere of copper collides with a rigid extrusion cone (shown in grey), producing a jet of ejected metal. The metal is surrounded by vacuum, using the void model outlined by \citet{WallisFluxEnriched}. The times are shown in $\mu$s, and the images depict the $y$-velocity and plastic strain. The results agree well with the experimental studies of \citet{GrayImpactExtrusion}.}
\label{fig:ImpactExtrusion}
\end{figure}

The set-up for the test is shown in Figure \ref{fig:ImpactExtrusionInitialConditions}. The copper is given an initial downward velocity of 400 ms$^{-1}$ and is governed by the equation of state equation \eqref{eq:Romenskii}, with parameters given in Table \ref{tab:Romenskii}. The test is run using the cylindrically symmetric form, with a reflective boundary condition along the $r=0$ axis. The domain spans $r=[0:5]$ mm, $z=[-40:22.5]$ mm. A base resolution of 56 $\times$ 696 is used, with two layers of AMR, each of refinement factor 2. The test is run until $t=60 \ \mu$s using a CFL of 0.4, and is shown in Figure \ref{fig:ImpactExtrusion}.

The results agree well with the experimental results \cite{GrayImpactExtrusion}, shown in Figure \ref{fig:ImpactExtrusionExperiment}, despite the lack of a material damage model. Additionally, the maximum plastic strain of 9.327 agrees well with the value of 9.3 quoted by \citet{UdaykumarGroove} and the range 6-9 quoted by \citet{GrayImpactExtrusion}. This test also serves to demonstrate the application of more complex plasticity models, such as the Johnson Cook model employed here, correctly producing necking effects in the solid jet. However, the choice of plasticity model and parameters has a significant effect on the relative fragment sizes, and a more in-depth comparison of different models would be needed to be able to draw further experimental comparison.

\begin{table*}
\begin{center}
\begin{tabular}{|c|c|c|c|c|c|c|c|c|}
\hline
Material & $\rho_0$ /kgm$^{-3}$ & ${\cal A}$ / 10$^{11}$ Pa & ${\cal B}$ / 10$^{11}$ Pa & ${\cal R}_1$ & ${\cal R}_2$ & $\Gamma$ & $C^V$ /Jkg$^{-1}$K$^{-1}$& $Q$ /MJkg$^{-1}$ \\
\hline
Reactant & 1905 & 778.1 & -0.0503 & 11.3 & 1.13 & 0.8938 & 1305.5 & 0 \\
Product  & -    & 14.81 & 0.6379 & 6.2 & 2.2 & 0.5 & 524.9 & 3.94 \\
\hline
\end{tabular} 
\caption{The JWL equation of state parameters for LX-17.}
\label{tab:JWL}

\begin{tabular}{|c|c|c|c|c|c|c|c|c|c|c|c|}
\hline
$a$ & $b$ & $c$ & $d$ & $e$ & $g$ & $x$ & $y$ & $z$ & $F_{ig}$ & $F_{G_1}$ & $F_{G_2}$ \\
\hline
0.22 & 0.667 & 0.667 & 1 & 0.667 & 0.667 & 7 & 3 & 1 & 0.02 & 0.8 & 0.8 \\
\hline
\end{tabular}~\\[0.1cm]
\begin{tabular}{|c|c|c|}
\hline
I /s$^{-1}$ & G$_1$ / ($10^{11}$ Pa)$^{-y}$ s$^{-1}$   & G$_2$ / ($10^{11}$ Pa)$^{-z}$ s$^{-1}$ \\
\hline
$4\times 10^{12}$& 4500 $\times 10^{6}$ & 30 $\times 10^{6}$ \\
\hline
\end{tabular}
\caption{The ignition and growth reaction rate parameters for LX-17.}
\label{tab:IandG}
\end{center}
\end{table*}

\begin{figure}
\centering
\includegraphics[width = 0.9\textwidth]{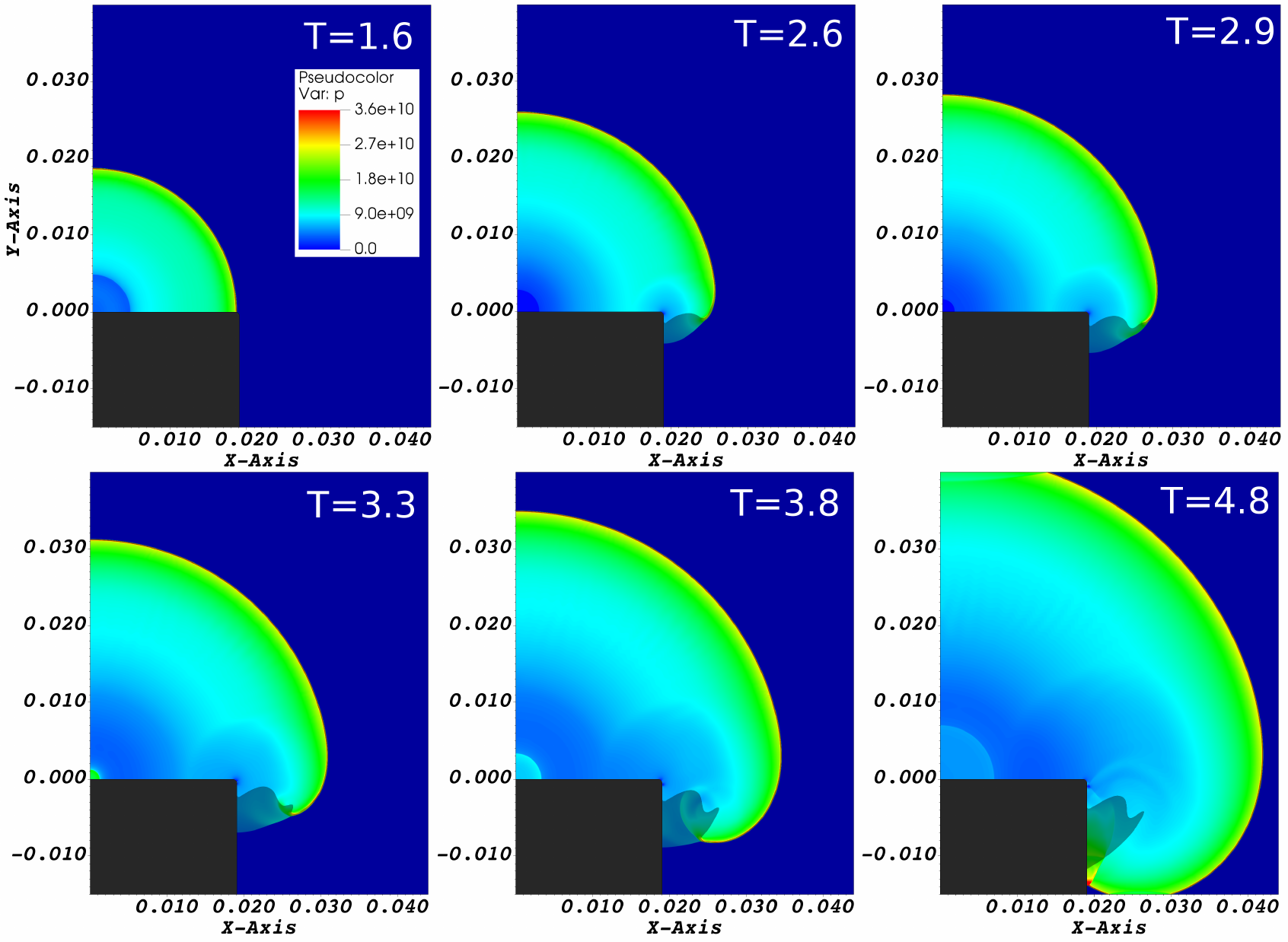}
\caption{The rigid LX-17 corner turning test, depicting the pressure in the explosive. Times are shown in $\mu$s. The dead-zones are tracked by overlaying a plot of the reaction progress variable with varying opacity; fully reacted material is transparent and unreacted material is slightly opaque.}
\label{fig:HockeyPuck_rigid}
\includegraphics[width = 0.9\textwidth]{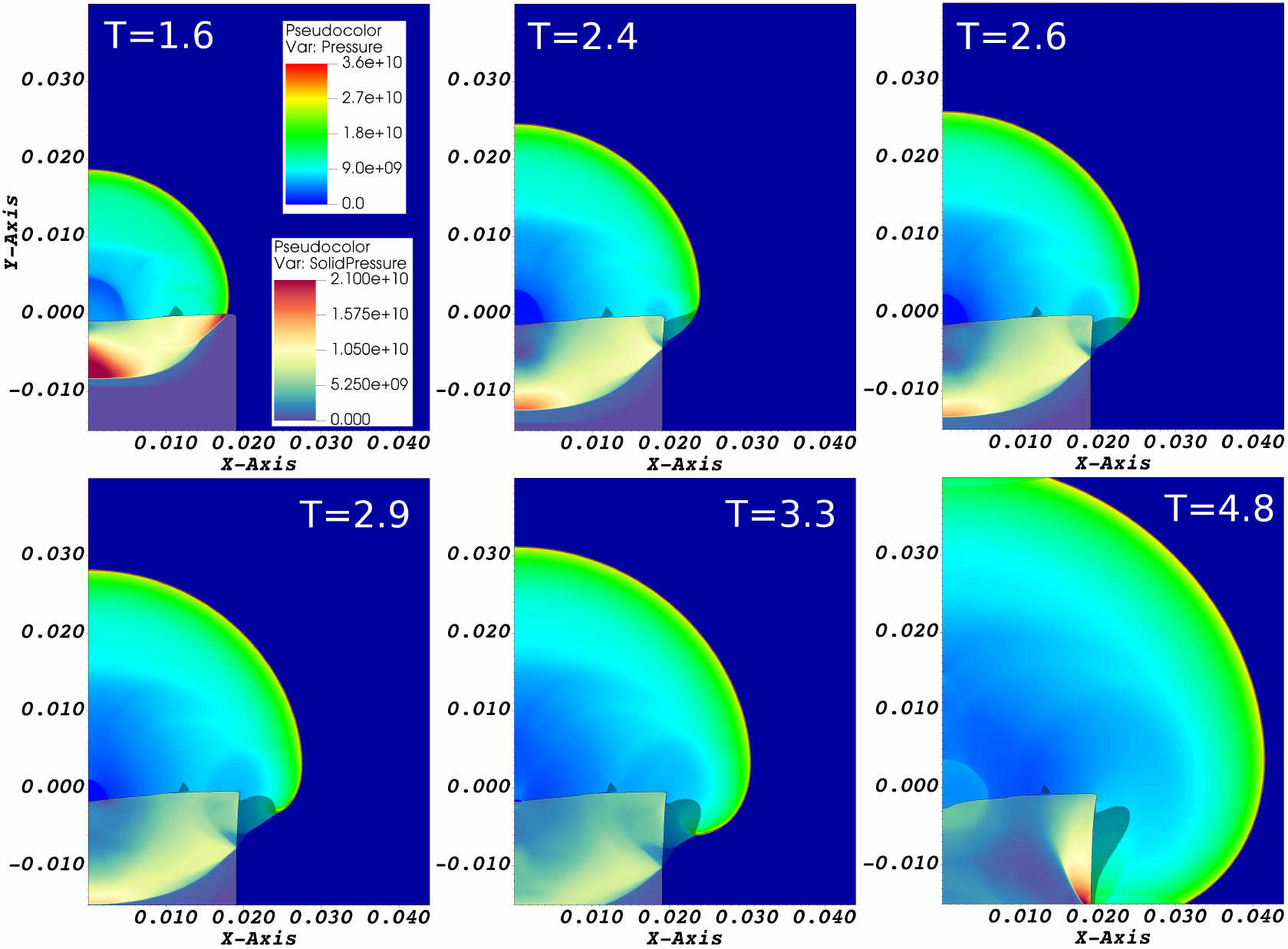}
\caption{The elastoplastic solid LX-17 corner turning test, depicting the pressure in both the explosive and the solid. Times are shown in $\mu$s. The dead-zones are tracked by overlaying a plot of the reaction progress variable with varying opacity; fully reacted material is transparent and unreacted material is slightly opaque.}
\label{fig:HockeyPuck_solid}
\end{figure}

\begin{figure}
\centering
\includegraphics[width = \textwidth]{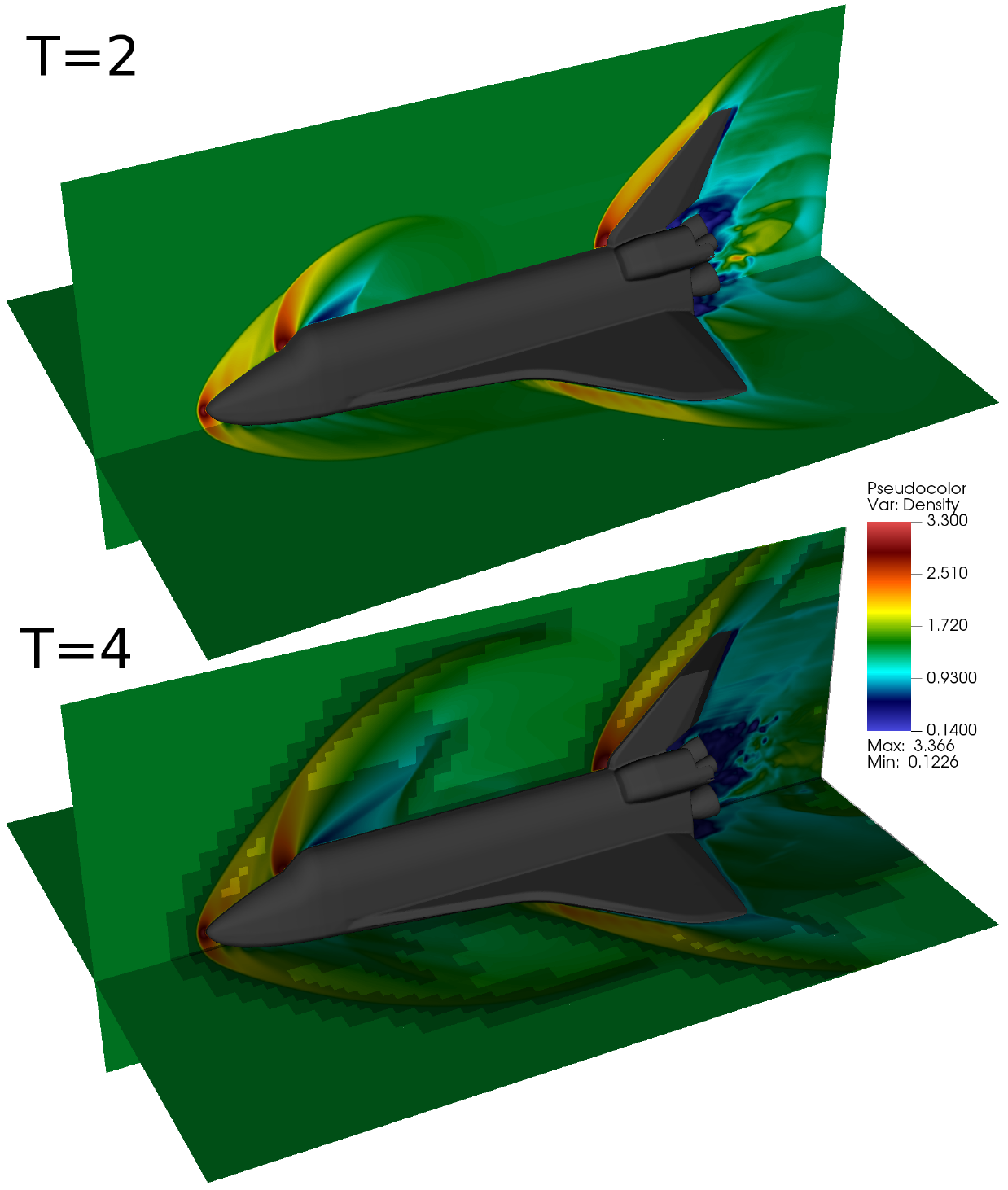}
\caption{The 3D STL geometry test. The test features a Mach 1.5 flow over a complex three-dimensional geometry, defined purely as volume fraction field. The images show the density and AMR levels at $t =2,4$. The method handles this difficult test well, and is straightforward to parallelise, even in three dimensions with AMR.}
\label{fig:Shuttle}
\end{figure}

\subsection{LX-17 Corner Turning}

To demonstrate the multi-physics capabilities of the current model, the desensitised LX-17 corner turning tests studied by \citet{SouersLX17CornerTurning}, \citet{DeoliveiraLX17CornerTurning} and \citet{TarverLX17CornerTurning} are considered. In this test, the domain is filled with the explosive LX-17, with a `hockey-puck' of solid or rigid material in the centre. A booster region directly above the puck ignites the explosive, and the subsequent detonation wave travels around the corner of the puck. There are several multi-physics aspects to this test. Firstly, the explosive LX-17 is modelled using the multi-phase formulation detailed by \citet{WallisMultiPhysics}, where both the reactants and products are explicitly modelled as a mixture with distinct equations of state. This mixture is then treated as one material for the purposes of bookkeeping, and the reaction progress variable $\lambda$ is added to the full multi-material system of equations in order to track the reaction:
\begin{align}
\frac{\partial \phi_{(l)}\rho_{(l)} \lambda_{(l)}}{\partial t} + \frac{\partial \phi_{(l)} \rho_{(l)}  \lambda_{(l)} u_k}{\partial x_k} &= \phi_{(l)}\rho_{(l)}\dot{\lambda}_{(l)} \ .
\end{align}
The explosive's products and reactants are governed by the JWL equation-of-state, with parameters given in Table \ref{tab:JWL}. This approach further allows for the use of the ignition-and-growth model, as employed by \citet{DeoliveiraLX17CornerTurning}, to simulate the rate of reaction of the explosive. This is a three-stage rate law, based on phenomenological experience of how detonations evolve in condensed phase explosives. The rate can be expressed as:
\begin{align}
\dot{\lambda} =& I(1-F)^b\left(\frac{\rho}{\rho_0}-1-a\right)^xH(F_{ig}-F) \\
               &+ G_1(1-F)^cF^dp^yH(F_{G_1}-F) \nonumber \\
               &+ G_2(1-F)^eF^gp^zH(F-F_{G_2}) \ , \nonumber
\end{align}
where $F = 1-\lambda$ is the reacted fraction and $H$ is the Heaviside function. Other parameters are material dependent constants, given in Table \ref{tab:IandG}. Full details of the method are given by \citet{WallisMultiPhysics}.

Secondly, an explosive desensitisation model is also implemented. The model was used by \citet{DeoliveiraLX17CornerTurning} to study the formation of `dead-zones' around the corner of the puck. These are regions where the explosive is not ignited as the detonation diffracts around the corner, due to the shock-compression of the material. This model is implemented by including an additional history variable $\varphi$ which tracks the desensitisation of the explosive:
\begin{align}
\frac{\partial \phi_{(l)}\rho_{(l)}\varphi_{(l)}}{\partial t} + \frac{\partial \phi_{(l)}\rho_{(l)} \varphi_{(l)} u_k}{\partial x_k} = \phi_{(l)}\rho_{(l)}Ap(1-\varphi_{(l)})(\varphi_{(l)}+\epsilon) \ ,
\end{align}
where $A=8.36 \times 10^{-3}$ is a material dependent constant, $p$ is the pressure and $\epsilon=10^{-3}$ is a small value designed to increase $\varphi$ from zero. When the explosive is shocked, the desensitisation goes from 0 (no desensitisation) to 1 (fully desensitised). This parameter is then fed into the ignition-and-growth reaction model to slow the reaction in areas that have been desensitised. To this end, parameter $a$ in the ignition-and-growth model is modified to:
\begin{align}
a(\varphi) = a_0(1-\varphi)+a_1\varphi \ ,
\end{align}
where $a_0$ was the original parameter, and $a_1=0.5$ is a parameter used to calibrate the desensitisation. Additionally, the first growth term, $G_1$, is now only activated when the reaction progress variable is greater than a threshold, $F_{G_1,\text{min}} = \lambda_c\varphi$, with $\lambda_c=0.01$ being another calibration parameter. This means the reaction rate is now:
\begin{align}
\dot{\lambda} =& I(1-F)^b\left(\frac{\rho}{\rho_0}-1-(a_0(1-\varphi)+a_1\varphi)\right)^xH(F_{ig}-F) \\
               &+ G_1(1-F)^cF^dp^yH(F_{G_1}-F)H(F-\lambda_c\varphi) \nonumber \\
               &+ G_2(1-F)^eF^gp^zH(F-F_{G_2}) \ . \nonumber
\end{align}

Thirdly, this test can either be performed with a rigid body or with an elastoplastic solid, demonstrating how the methodology at hand is capable of transitioning between either approach with no drastic change of numerical scheme. The solid in this case is steel, governed by the equation of state equation \eqref{eq:Romenskii} with the parameters given in Table \ref{tab:Romenskii}, obeying a ideal plasticity law with a yield stress of $\sigma_Y=1.37$ GPa.

Both of tests have the same set-up. The tests are run in the cylindrically symmetric formulation, adding radial source terms to model the effect of the geometry and including a reflective boundary condition at the $r=0$ axis. The domain spans $r=[0:44]$ mm, $z=[-15:40]$ mm. A base resolution of $272 \times 344$ is employed, with 3 levels of AMR each of refinement factor 2. The tests are run to a final time of 5 $\mu$s with a CFL of 0.4. 

The puck has a height of 15 mm, a radius of 19.05 mm and a rounded corner with a radius of 0.5 mm. The detonation is ignited by a hemispherical booster region on top of the puck, centred at (0,0) with a radius of 7.68 mm.  

Both the rigid and elastoplastic tests are shown in Figure \ref{fig:HockeyPuck_rigid} and \ref{fig:HockeyPuck_solid}. The tests depict the pressure profile over the course of the test, showing how the detonation diffracts around the corner. The dead-zones are tracked by overlaying a plot of the reaction progress variable with varying opacity; fully reacted material is transparent and unreacted material is slightly opaque. Both cases handle the test well and demonstrate the marked difference between discretising inside the solid, compared to assuming a rigid body. The rigid tests agree well with the results of \citet{DeoliveiraLX17CornerTurning}.

\subsection{Stereolithography (STL) Geometry}

The method is designed to be easily extendible to large 3D simulations with multiple levels of AMR. This example is designed to test the ability of the model to handle arbitrary three-dimensional geometries, such as would be used in research applications. In the test, the geometry in question is a scale model of the NASA space shuttle \cite{NasaShuttle}, manually edited to form a watertight geometry. The volume fraction field of the body is extracted from the triangulated surface mesh by first using the VTK \cite{VTK} object \texttt{vtkImplicitPolyDataDistance} to obtain a signed distance function. From here, the volume of the body in a cell is computed as the volume of the polyhedron formed by the intersection of the signed distance function zero contour with the cell, using the formula from \citet{BartonLevelSet}. The volume fraction is then set to be the rigid body volume fraction.  

The test is non-dimensionalised, with a Mach 1.5 flow over the body initialised as $\rho=1.4, p=1, \vb{u}=1.5\hat{x}$. The domain spans $x=[-10.0:10.0]$, $y=[-7.5:7.5]$, $z=[-5.0:5.0]$, with a base resolution of $176\times128\times88$ and 2 levels of AMR, each of refinement factor 2. The test is run until $t=4$, using a CFL of 0.3.

The test is shown in Figure \ref{fig:Shuttle}, depicting the density and AMR levels at $t=2,4$. The test performs well, with the method working well even with complex three-dimensional geometries.

\section{Conclusions}

This work has demonstrated a straightforward approach to the inclusion of immersed rigid boundaries in a variety of multi-physics applications. The example test cases have included single fluid tests with both static and dynamic rigid bodies, the interaction of elastoplastic solids with rigid bodies, multi-phase reactive fluids, and complex three-dimensional geometries. All of these test cases are modelled using the same diffuse interface model. The rigid body is included as a scalar field, and all material interaction is incorporated into the underlying system of equations which can handle an arbitrary number of solid and fluid components. The model is shown to be extendible to three dimensions and adaptive mesh refinement. This is thanks to the local, parallelisable nature of both the hyperbolic update and the interface seeding method which mediates the rigid body boundary conditions.

\section*{Acknowledgements}

This work was funded by AWE PLC. Additionally, Tim Wallis is supported by a grant from the UK Engineering and Physical Sciences Research Council (EPSRC) EP/L015552/1 for the Centre for Doctoral Training (CDT) in Computational Methods for Materials Science.

\section*{Author declaration}

The authors have no conflicts of interest to disclose.

\section*{Data availability statement}

The data that support the findings of this study are available within the article. 

\section*{References}

\bibliography{Rigidbib}

\end{document}